\documentclass[a4paper,11pt]{article}
\usepackage{jcappub}

\usepackage{mathtools}
\usepackage{appendix}
\usepackage[T1]{fontenc}

\newcommand{\rR}{\rho_R}
\newcommand{\rp}{\rho_\phi}
\newcommand{\Gp}{\Gamma_\phi}
\newcommand{\gs}{g_\star}
\newcommand{\gss}{g_{\star s}}
\newcommand{\Tfo}{T_\text{fo}}
\newcommand{\xfo}{x_\text{fo}}
\newcommand{\afo}{a_\text{fo}}
\newcommand{\arh}{a_\text{rh}}
\newcommand{\Trh}{T_\text{rh}}
\newcommand{\Hrh}{H_\text{rh}}

\newcommand{\sv}{\langle\sigma v\rangle}
\newcommand{\Odm}{\Omega_\text{dm}}
\newcommand{\ndm}{n_\text{dm}}

\title{WIMPs during Reheating}

\author[a,\, b]{Nicolás Bernal}
\author[c]{and Yong Xu}

\affiliation[a]{New York University Abu Dhabi\\
PO Box 129188, Saadiyat Island, Abu Dhabi, United Arab Emirates}
\affiliation[b]{Centro de Investigaciones, Universidad Antonio Nariño\\
Carrera 3 este \# 47A-15, Bogotá, Colombia}
\affiliation[c]{Bethe Center for Theoretical Physics and Physikalisches Institut, Universität Bonn\\
Nussallee 12, 53115 Bonn, Germany}

\emailAdd{nicolas.bernal@nyu.edu}
\emailAdd{yongxu@th.physik.uni-bonn.de}

\abstract{Weakly Interacting Massive Particles (WIMPs) are among the best-motivated dark matter candidates. In the standard scenario where the freeze-out occurs well after the end of inflationary reheating, they are in tension with the severe experimental constraints. Here, we investigate the thermal freeze-out of WIMPs occurring {\it during} reheating, while the inflaton $\phi$ coherently oscillates in a generic potential $\propto \phi^n$. Depending on the value of $n$ and the spin of the inflaton decaying products, the evolution of the radiation and inflaton energy densities can show distinct features, therefore, having a considerable impact on the freeze-out behavior of WIMPs. As a result of the injection of entropy during reheating, the parameter space compatible with the observed DM relic abundance is enlarged. In particular, the WIMP thermally averaged annihilation cross-section can be several magnitudes lower than that in the standard case. Finally, we discuss the current bounds from dark matter indirect detection experiments, and explore future challenges and opportunities.}
\begin{document}
\begin{flushright}
    PI/UAN-2022-722FT
\end{flushright}
\maketitle

\section{Introduction}
The existence of non-baryonic dark matter (DM) in the Universe is very compelling, as suggested by both astrophysical and cosmological observations~\cite{Bertone:2016nfn}. A viable DM candidate has to satisfy several properties: It has to be electromagnetically neutral, stable at the cosmological scales, and non-relativistic at the matter-radiation equality in order to allow structure formation. Finally, it must feature a relic density $\Odm h^2 \simeq 0.12$, accounting for 27\% of the total energy budget of the Universe~\cite{Planck:2018vyg, Drees:2018hzm}.

The most prominent DM production mechanism in the early universe corresponds to the weakly interacting massive particle (WIMP) paradigm~\cite{Arcadi:2017kky, Roszkowski:2017nbc}.
In the standard WIMP scenario, DM has a mass at the electroweak scale and couples to the standard model (SM) thermal plasma with a sizable strength, typical of electroweak interactions.
WIMPs reach thermal equilibrium with the SM thermal plasma and eventually {\it freeze out}, giving rise to the observed DM relic abundance.
This DM freeze-out is commonly assumed to occur well after the end of reheating, when the Universe energy density is dominated by SM radiation.
To match observations, a thermally averaged annihilation cross-section $\sv = \text{few} \times 10^{-26}$~cm$^3$/s is typically required~\cite{Steigman:2012nb}.
The WIMP mechanism is particularly interesting because it could be tested in a number of complementary ways, including direct, indirect, and collider probes.
However, current null experimental results and severe constraints on the natural parameter space motivate some quests beyond the standard WIMP paradigm~\cite{Arcadi:2017kky}.

There are several possibilities to evade the strong experimental constraints. For example, DM might have feeble interactions with the SM thermal plasma so that it is produced out of equilibrium by the so-called freeze-in mechanism (FIMP)~\cite{McDonald:2001vt, Choi:2005vq, Kusenko:2006rh, Petraki:2007gq, Hall:2009bx, Elahi:2014fsa, Bernal:2017kxu}. Alternatively, WIMPs might have decoupled during a non-standard cosmological epoch~\cite{Allahverdi:2020bys}, with an annihilation cross section much smaller than the one enforced in the standard case~\cite{Arias:2019uol}.
In general, the production of DM in scenarios with a non-standard expansion phase has recently gained increasing interest; see, e.g., Refs.~\cite{Kane:2015jia, Co:2015pka, Davoudiasl:2015vba, Randall:2015xza, Berlin:2016vnh, Tenkanen:2016jic, Dror:2016rxc, Berlin:2016gtr, DEramo:2017gpl, Hamdan:2017psw, Visinelli:2017qga, Dror:2017gjq, Drees:2017iod, DEramo:2017ecx, Maity:2018dgy, Bernal:2018ins, Hardy:2018bph, Maity:2018exj, Hambye:2018qjv, Bernal:2018kcw, Arbey:2018uho, Drees:2018dsj, Betancur:2018xtj, Maldonado:2019qmp, Poulin:2019omz, Tenkanen:2019cik, Arias:2019uol, Han:2019vxi, Bernal:2019mhf, Chanda:2019xyl, Mahanta:2019sfo, Arcadi:2020aot, Konar:2020vuu, Bhatia:2020itt, Arias:2020qty, Barman:2021ifu, Arcadi:2021doo, Mahanta:2022gsi}.
For earlier work, see also Refs.~\cite{Barrow:1982ei, Kamionkowski:1990ni, McDonald:1989jd, Salati:2002md, Comelli:2003cv, Rosati:2003yw, Pallis:2004yy, Gelmini:2006pw, Gelmini:2006pq, Arbey:2008kv, Cohen:2008nb, Arbey:2009gt}.
Moreover, scenarios where freeze-out occurs during reheating are also viable, and in such cases, the freeze-out temperature $\Tfo$ is higher than the reheating temperature $\Trh$; note that the latter characterizes the onset of the radiation-dominated era.
We note that WIMP scenarios with low reheating temperature have been widely investigated in the literature~\cite{Giudice:2000ex, Fornengo:2002db, Gelmini:2006pw, Drees:2006vh, Roszkowski:2014lga, Drees:2017iod}, usually triggered by the decay of a long-lived scalar field with an equation-of-state (EoS) parameter $\omega = 0$ and a constant decay rate $\Gp$.\footnote{For studies on baryogenesis with a low reheating temperature or during an early matter-dominated phase, see Refs.~\cite{Davidson:2000dw, Giudice:2000ex, Allahverdi:2010im, Beniwal:2017eik, Allahverdi:2017edd} and~\cite{Bernal:2017zvx, Chen:2019etb, Chakraborty:2022gob}, respectively. Furthermore, primordial gravitational wave production in scenarios with an early matter era has recently received particular attention~\cite{Assadullahi:2009nf, Durrer:2011bi, Alabidi:2013lya, DEramo:2019tit, Bernal:2019lpc, Figueroa:2019paj, Bernal:2020ywq}.}
Finally, we note that this kind of scenario in which DM reaches chemical equilibrium with the SM bath eliminates the possible overproduction of DM during inflation~\cite{Garcia:2022vwm, Kaneta:2022gug}.

In this work, different from the aforementioned literature, we investigate the phenomenology of WIMP freezing out during reheating, where the inflaton field $\phi$ oscillates around the minimum of a generic potential $\propto \phi^n$.
Such potentials naturally arise in a number of inflationary scenarios like the $\alpha$-attractor $T$-model~\cite{Kallosh:2013hoa, Kallosh:2013yoa}, or the Starobinsky model~\cite{Starobinsky:1980te, Starobinsky:1981vz, Starobinsky:1983zz, Kofman:1985aw}.
In such a case, both the inflaton EoS and its decay rate are modified compared to the usually assumed case (i.e. $n = 2$). In particular, the EoS depends on the shape of the potential $\omega = (n-2)/(n+2)$~\cite{Turner:1983he}, and $\Gp$ develops a time dependency~\cite{Garcia:2020eof, Garcia:2020wiy}. As a consequence, the inflaton and SM energy densities exhibit a non-standard evolution during reheating~\cite{Co:2020xaf}, which has a strong impact on WIMP decoupling.\footnote{We note that the phenomenology of FIMPs~\cite{Garcia:2020eof, Garcia:2020wiy, Ahmed:2021fvt, Barman:2022tzk, Ahmed:2022tfm} and the QCD axion DM~\cite{Arias:2022qjt} with a time-dependent decay rate has recently been analyzed.} Depending on the value of $n$, the inflaton decay products, as well as the reheating temperature, we find that due to the large injection of entropy during reheating, $\sv$ can be several orders of magnitude smaller than in the standard case, thus evading the experimental constraints. We highlight that some regions of the new parameter space are currently being tested by several experiments. Furthermore, we notice that the future CTA experiment might increase our ability to probe different scenarios.

The remainder of the paper is organized as follows. In Sect.~\ref{sec:reheating} we revisit inflationary reheating, in particular, the evolution of the inflaton and the radiation energy densities, with an inflaton oscillating in a potential $\propto \phi^n$. In Sect.~\ref{sec:DM_production}, we investigate WIMP DM production focusing on the case where freeze-out occurs {\it during} reheating. Our findings are summarized in Sect.~\ref{sec:concl}.

\section{Cosmology during Reheating} \label{sec:reheating}
We consider an inflaton $\phi$ that, after the end of inflation, oscillates at the bottom of a monomial potential $V$ of the form
\begin{equation}
    V(\phi) = \lambda\, \frac{\phi^n}{\Lambda^{n - 4}}\,,
\end{equation}
where $\lambda$ is a dimensionless coupling and $\Lambda$ an energy scale.
Potentials with these types of minima naturally arise in a number of inflationary scenarios, such as the $\alpha$-attractor $T$-model~\cite{Kallosh:2013hoa, Kallosh:2013yoa}, or the Starobinsky model~\cite{Starobinsky:1980te, Starobinsky:1981vz, Starobinsky:1983zz, Kofman:1985aw}.
The equation of motion for the oscillating inflaton field is given by~\cite{Turner:1983he}
\begin{equation} \label{eq:eom0}
    \ddot\phi + (3\, H + \Gp)\, \dot\phi + V'(\phi) = 0\,,
\end{equation} 
where $H$ denotes the Hubble expansion rate and $\Gp$ the inflaton decay rate.
The dots and primes correspond to derivatives with respect to time $t$ and $\phi$, respectively.
By multiplying the above expression by $\dot\phi$, Eq.~\eqref{eq:eom0} can be rewritten as
\begin{equation} \label{eq:eom}
    \frac{d}{dt} \left[\frac12 \dot\phi^2 + V(\phi)\right]= -(3\, H + \Gp)\, \dot\phi^2\,.
\end{equation}
By defining the energy density and pressure of $\phi$ as $\rp \equiv \frac12\, \dot\phi^2+ V(\phi)$ and $p_\phi \equiv \frac12\, \dot\phi^2 - V(\phi)$, together with the equation-of-state parameter $\omega \equiv p_\phi/\rp = (n - 2) / (n + 2)$~\cite{Turner:1983he},
Eq.~\eqref{eq:eom} gives the evolution of $\rp$ as
\begin{equation} \label{eq:drhodt}
    \frac{d\rp}{dt} + \frac{6\, n}{2 + n}\, H\, \rp = - \frac{2\, n}{2 + n}\, \Gp\, \rp\,.
\end{equation}
During reheating, that is, in the range $a_I \ll a \ll \arh$, where $a$ is the scale factor, and $a_I$ and $\arh$ correspond to the scale factors at the beginning and end of the reheating, respectively, Eq.~\eqref{eq:drhodt} admits the analytical solution
\begin{equation} \label{eq:rpsol}
    \rp(a) \simeq \rp (\arh) \left(\frac{\arh}{a}\right)^\frac{6\, n}{2 + n}.
\end{equation}
As during reheating the Hubble expansion rate is dominated by the inflaton energy density, it follows that
\begin{equation} \label{eq:Hubble}
    H(a) \simeq
    \begin{dcases}
        \Hrh \left(\frac{\arh}{a}\right)^\frac{3\, n}{n + 2} &\text{ for } a \leq \arh\,,\\
        \Hrh \left(\frac{\arh}{a}\right)^2 &\text{ for } \arh \leq a\,.
    \end{dcases}
\end{equation}
At the end of reheating (i.e., at $a = \arh$), the inflaton and radiation energy densities are equal, $\rR(\arh) = \rp(\arh) = 3\, M_P^2\, \Hrh^2$.
Note that in order not to spoil BBN, the reheating temperature must satisfy $\Trh\geq T_\text{BBN} \simeq 4$~MeV~\cite{Sarkar:1995dd, Kawasaki:2000en, Hannestad:2004px, DeBernardis:2008zz, deSalas:2015glj}.

During reheating, the inflaton transfers its energy density to SM radiation, the end of reheating corresponding to the onset of the SM radiation-dominated era.
The SM radiation energy density $\rR$ is governed by the Boltzmann equation~\cite{Garcia:2020wiy}
\begin{equation} \label{eq:rR}
    \frac{d\rR}{dt} + 4\, H\, \rR = + \frac{2\, n}{2 + n}\, \Gp\, \rp\,.
\end{equation}
Using Eq.~\eqref{eq:rpsol}, one can solve Eq.~\eqref{eq:rR} and further obtain the solution for the evolution of the radiation energy density during reheating 
\begin{equation} \label{eq:rR_int}
    \rR(a) \simeq \frac{2\, \sqrt{3}\, n}{2 + n}\, \frac{M_P}{a^4} \int_{a_I}^a \Gp(a')\, \sqrt{\rp(a')}\, a'^3\, da',
\end{equation}
where a general scale factor dependence of $\Gp$ has been assumed.
Such a dependence can come, for example, from the inflaton mass parameter.
The effective mass $m_\phi$ for the inflaton field, understood as the second derivative of its potential, is given by
\begin{equation}
    m_\phi^2 \equiv \frac{d^2V}{d\phi^2} = n\, (n - 1)\, \lambda\, \frac{\phi^{n - 2}}{\Lambda^{n - 4}}
    \simeq n\, (n-1)\, \lambda^\frac{2}{n}\, \Lambda^\frac{2\, (4 - n)}{n} \rp(a)^{\frac{n-2}{n}}. 
\end{equation}
It is interesting to note that for $n \neq 2$, $m_\phi$ features a field dependence that, in turn, would lead to an inflaton decay rate with a scale factor (or time) dependence.

In the following, two different reheating scenarios will be investigated in which the inflaton either decays to a pair of fermions or bosons via a trilinear coupling.

\subsection{Fermionic Reheating}
In the case where the inflaton only decays into a pair of fermions $\Psi$ and $\bar\Psi$ via a trilinear interaction $y\, \phi\, \bar\Psi\, \Psi$, with $y$ being the corresponding Yukawa coupling, the decay rate is given by
\begin{equation} \label{eq:fer_gamma}
    \Gp = \frac{y_{\text{eff}}^2}{8\pi}\, m_\phi\,,
\end{equation}
where the effective coupling $y_\text{eff} \ne y$ (for $n\neq2$) is obtained after averaging over oscillations~\cite{Shtanov:1994ce, Garcia:2020wiy, Ichikawa:2008ne}. We will return to our treatment for effective coupling shortly.
The evolution of the SM energy density in Eq.~\eqref{eq:rR_int} becomes
\begin{equation} \label{eq:rR_sol}
    \rR(a) \simeq \frac{\sqrt{3}}{8 \pi}\, \frac{n \sqrt{n (n - 1)}}{7 - n}\, \frac{y_\text{eff}^2\, \lambda^\frac{1}{n}\, M_P\, \rp(\arh)^\frac{n - 1}{n}}{\Lambda^\frac{n - 4}{n}} \left(\frac{\arh}{a}\right)^\frac{6 (n - 1)}{2 + n} \left[1 - \left(\frac{a_I}{a}\right)^\frac{2 (7 - n)}{2 + n}\right],
\end{equation}
and, therefore, the temperature of the SM bath evolves as
\begin{equation} \label{eq:Tfer}
    T(a) \simeq \Trh \times
    \begin{dcases}
        \left(\frac{\arh}{a}\right)^{\frac32 \frac{n - 1}{n + 2}} &\text{ for } n < 7\,,\\
        \frac{\arh}{a} &\text{ for } n > 7\,.
    \end{dcases}
\end{equation}
The Hubble expansion rate (cf. Eq.~\eqref{eq:Hubble}) can be rewritten as a function of $T$ as
\begin{equation}
    H(T) \simeq \Hrh
    \begin{dcases}
        \left(\frac{T}{\Trh}\right)^\frac{2\, n}{n - 1} &\text{ for } n < 7\,,\\
        \left(\frac{T}{\Trh}\right)^\frac{3\, n}{n + 2} &\text{ for } n > 7\,.
    \end{dcases}
\end{equation}
Before closing this subsection, we note that for the case with $n = 2$ where the inflaton oscillates in a quadratic potential with an EoS parameter $\omega=0$, the standard dependences with the scale factor $\rR(a) \propto a^{-3/2} $ and $T(a) \propto a^{-3/8}$ are reproduced.

\subsection{Bosonic Reheating}
Alternatively, if the inflaton only decays into a pair of bosons $S$ through the interaction $\mu\, \phi\, S\, S$, with $\mu$ being a coupling with mass dimension, the decay rate is instead
\begin{equation}
       \Gp = \frac{\mu_{\text{eff}}^2}{8\pi\, m_\phi}\,,
\end{equation}
where again the effective coupling $\mu_\text{eff} \ne \mu$ (if $n\neq2$) can be obtained after averaging over oscillations.
Using a procedure similar to the previous fermionic case, one sees that the SM energy density scales as
\begin{equation}
    \rR(a) \simeq \frac{\sqrt{3}}{8 \pi} \frac{1}{1 + 2\, n} \sqrt{\frac{n}{n - 1}}\, \frac{\mu_{\text{eff}}^2}{\lambda^{1/n}}\, M_P\, \Lambda^\frac{n - 4}{n}\, \rp(\arh)^\frac{1}{n} \left(\frac{\arh}{a}\right)^\frac{6}{2 + n} \left[1 - \left(\frac{a_I}{a}\right)^\frac{2\, (1 + 2 n)}{2 + n}\right],
\end{equation}
with which the SM temperature evolves as
\begin{equation}
    T(a) \simeq \Trh \label{eq:TBos} \left(\frac{\arh}{a}\right)^{\frac32 \frac{1}{2 + n}},
\end{equation}
and Hubble as
\begin{equation} \label{eq:Hubble_bos}
    H(T) \simeq \Hrh \left(\frac{T}{\Trh}\right)^{2\, n},
\end{equation}
during reheating.

\begin{figure}[t!]
    \def\sepf{0.51}
	\centering
    \includegraphics[scale=\sepf]{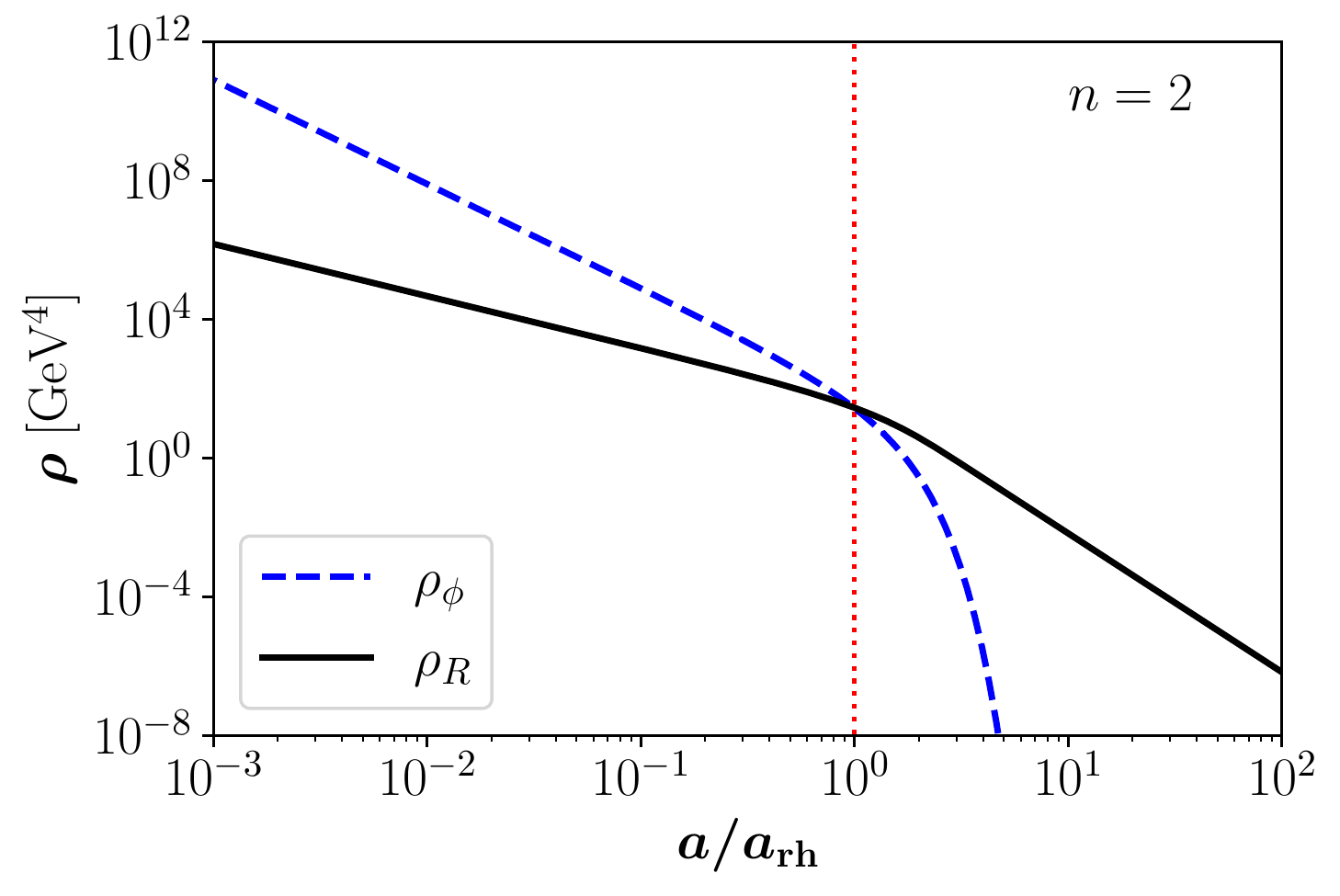}
    \includegraphics[scale=\sepf]{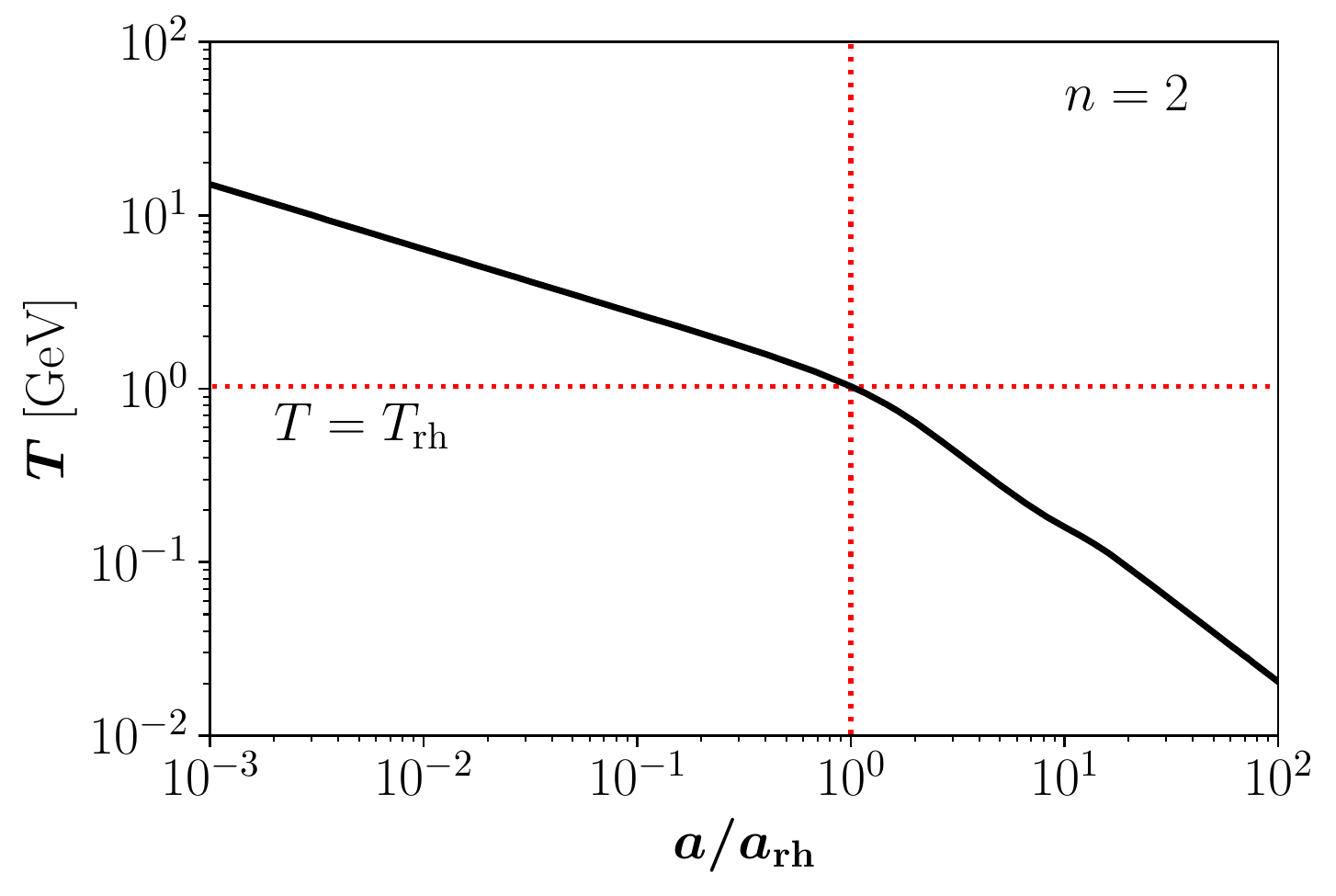}
    \includegraphics[scale=\sepf]{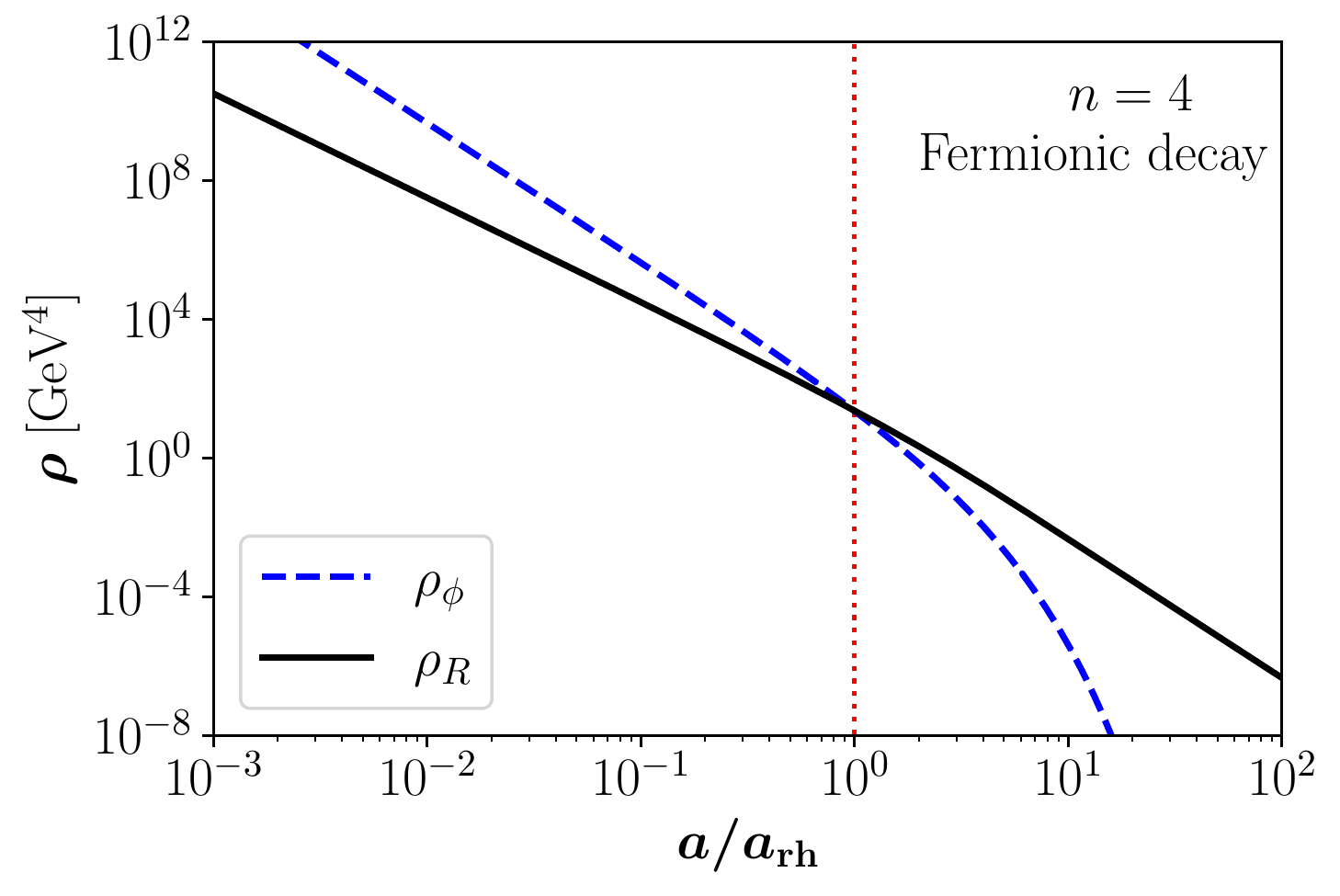}
    \includegraphics[scale=\sepf]{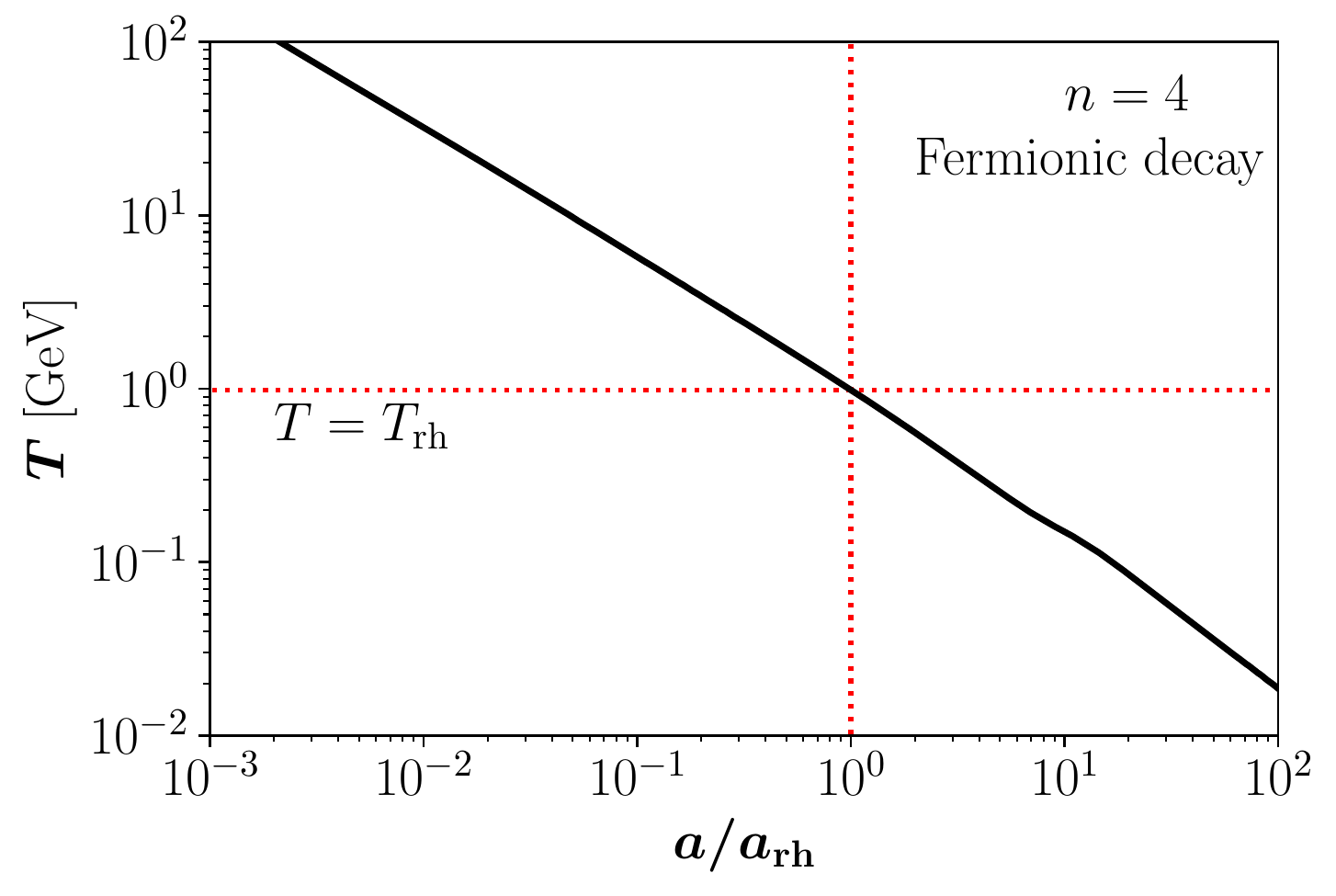}
    \includegraphics[scale=\sepf]{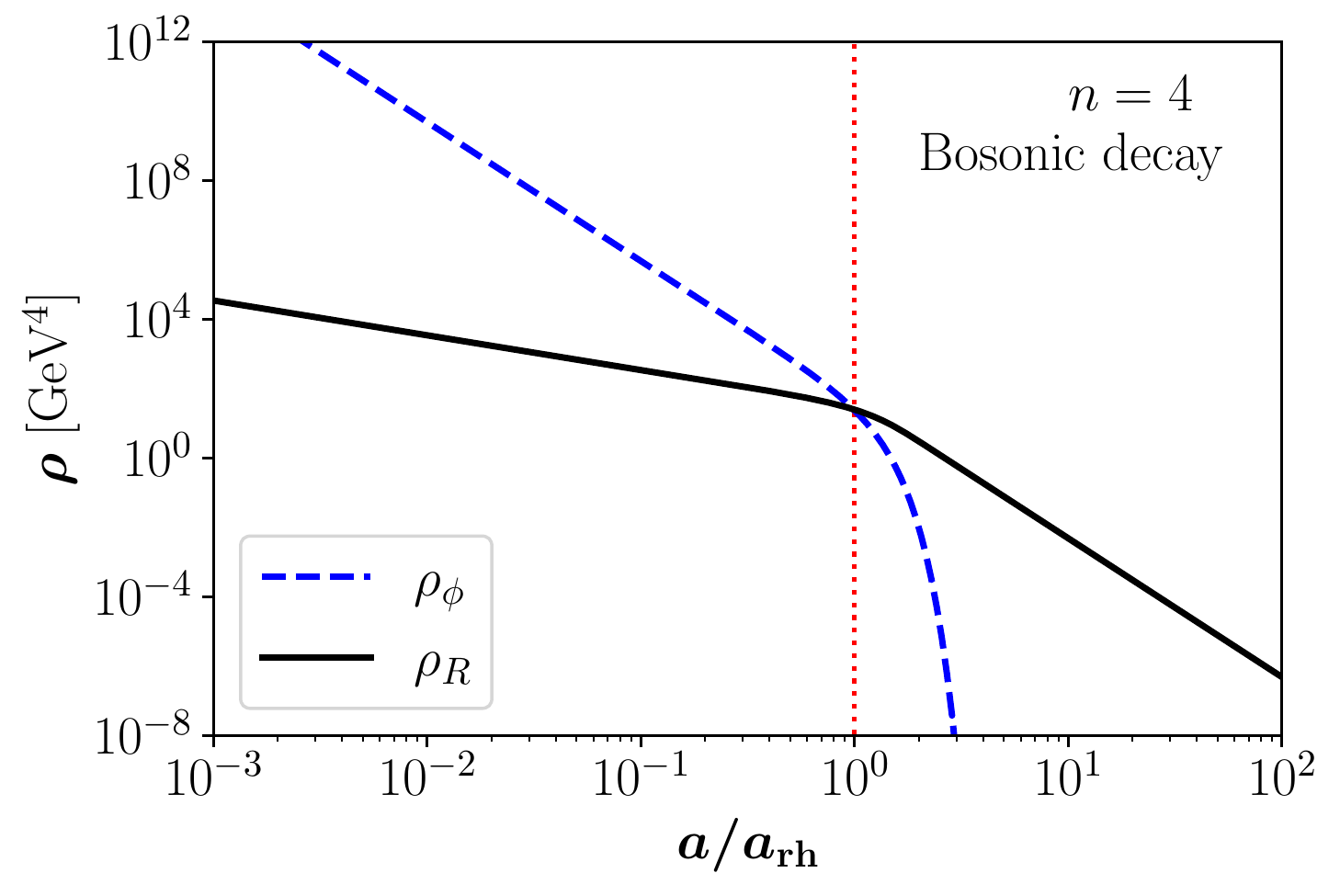}
    \includegraphics[scale=\sepf]{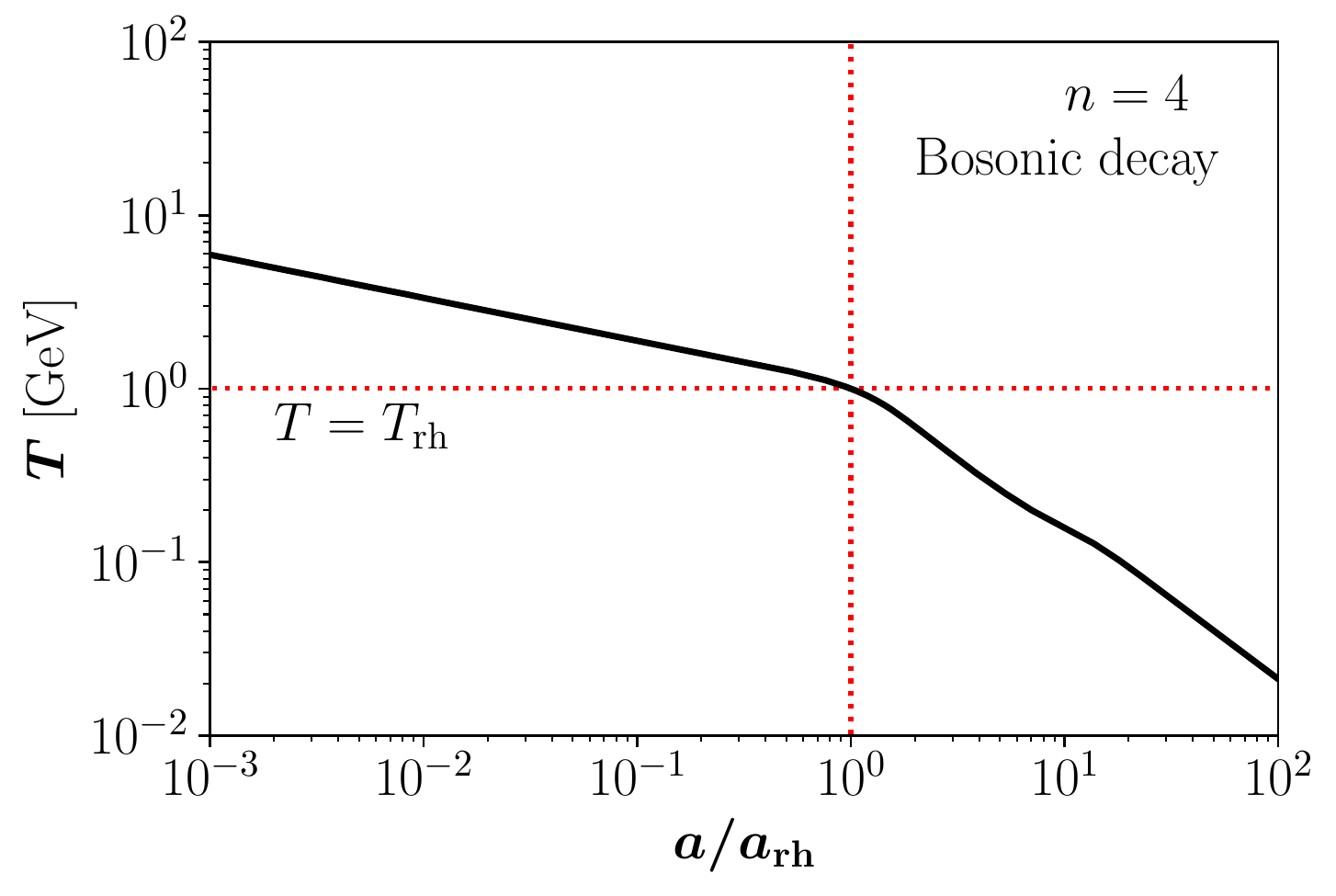}
    \caption{Left: Numerical evolution of the energy densities of the SM (solid black) and $\phi$ (dashed blue), as a function of the scale factor, for $\Trh = 1$~GeV.
    Right: Corresponding evolution of the SM temperature.
    }
	\label{fig:rho}
\end{figure} 
In Fig.~\ref{fig:rho}, the left panels depict the evolution of both the SM radiation (solid black) and the inflaton (dashed blue) energy densities, as a function of the scale factor, for $\Trh = 1$~GeV.
The corresponding evolution of the SM bath temperature is shown in the right panels. We note that the bump at $T \sim 0.1$~GeV is due to the QCD phase transition, where the latter leads to a sudden decrease of effective relativistic degrees of freedom in the thermal plasma.
The upper panels correspond to the case $n = 2$, where the effects of bosonic and fermionic decays are equivalent.
During reheating, $\rp \propto a^{-3}$, $\rR \propto a^{-3/2}$, and $T \propto a^{-3/8}$.
The middle and lower panels correspond to $n = 4$ with fermionic or bosonic decay, respectively.
For the first case, $\rR \propto a^{-3}$ (and $T \propto a^{-3/4}$) while for the latter case, $\rR \propto a^{-1}$ (and $T \propto a^{-1/4}$). Note that to numerically solve the Boltzmann equations for $\rp$ and $\rR$, the values of the effective couplings appearing in the decay rates are needed, which shall be obtained numerically~\cite{Garcia:2020wiy}. Instead, here we fix $\Trh$ and then calculate the effective coupling values so that the reheating temperature matches the desired value. Let us remind the reader that the reheating temperature has been defined as $\rR(\Trh) = \rp(\Trh) = 3\, M_P^2\, \Hrh^2$.

Before proceeding, a comment on the preheating effect is necessary. With the help of lattice simulations, it has been shown that, during reheating, the EoS parameter $\omega \to 1/3$ for $n \gtrsim 3$ due to inflaton fragmentation and parametric resonance~\cite{Lozanov:2016hid, Maity:2018qhi, Saha:2020bis, Antusch:2020iyq}. 
This implies that preheating efficiently drives the Universe to a radiation-dominated phase. However, to fully deplete the inflaton energy, perturbative decay through trilinear couplings between the inflaton and the daughter particles is still required, which is expected to occur in the last stage of the heating process after inflation~\cite{Maity:2018qhi}. Note that for efficient inflaton energy transfer via preheating, sizable couplings are usually required, which could spoil the inflaton potential and inflationary predictions due to loop corrections. Furthermore, in the literature, coherent oscillations of the inflaton condensate have been found to break down~\cite{Easther:2010mr}, which could delay preheating. It has also been shown that depending on the spin of the daughter particles, preheating might be shut down due to the large effective vacuum expectation value of the Higgs field~\cite{Freese:2017ace}. In this work, we are interested in low-reheating-temperature scenarios, so we assume small inflaton couplings to daughter particles, in which case preheating effects are not expected to be efficient. Thereafter, throughout this work, we will focus on the perturbative reheating scenario.

Having understood possible cosmological histories for the background, in the next section the evolution of the DM number density will be studied, taking particular care to the case where DM freezes out during reheating.

\section{Dark Matter Production} \label{sec:DM_production}
The evolution of the DM number density $\ndm$ is governed by the Boltzmann equation
\begin{equation} \label{eq:Bzeq1}
    \frac{d\ndm}{dt}+3\, H\, \ndm = - \sv \left(\ndm^2-n_{\mathrm{eq}}^{2}\right),
\end{equation}
with the Hubble expansion rate
\begin{equation} \label{eq:Htotal}
    H^2 = \frac{\rR + \rp}{3\, M_P^2}\,.
\end{equation}
$n_\text{eq}(T)$ corresponds to the equilibrium number density given by
\begin{equation}
    n_\text{eq}(T) \simeq g \left(\frac{m\,T}{2\pi} \right)^{3/2} e^{-\frac{m}{T}}\,
\end{equation}
for non-relativistic particles, where $m$ and $g$ denote the mass and the number of degrees of freedom of the DM field, respectively.
In this work, we focus on a temperature-independent cross section, which corresponds to an $S$-wave dominated interaction.
We note that direct production of DM out of the inflaton decay may also be relevant. In that case, Eq.~\eqref{eq:Bzeq1} would have an additional term $\propto 2\, \text{Br}\, \Gamma_\phi\, n_\phi$, where $\text{Br}$ denotes the branching ratio of the inflaton decay into a couple of DM particles, and $n_\phi$ is the inflaton number density.
Here we focus on the effect of the time dependence of $\Gp$ on the evolution of the WIMP thermal production, and therefore, in the following analysis this contribution will be omitted.%
\footnote{This is typically a good assumption as long as $\text{Br} \lesssim 10^{-4}\, m/(100~\text{GeV})$~\cite{Drees:2017iod, Arias:2019uol}.}
Finally, for a discussion regarding DM production during the SM thermalization process, we refer the reader to Refs.~\cite{Harigaya:2014waa, Harigaya:2019tzu, Drees:2021lbm, Drees:2022vvn, Mukaida:2022bbo}.

\subsection{Freeze out after Reheating}
We first revisit the standard scenario, where DM freeze-out occurs well {\it after} reheating, in the standard radiation-dominated era.
In this case, since the SM entropy is conserved during and after the DM freeze-out, Eq.~\eqref{eq:Bzeq1} can conveniently be rewritten as 
\begin{equation} \label{eq:Bzeq2}
    \frac{dY}{d\,x} = -\frac{\sv\, s}{H\, x} \left(Y^2 - Y^2_{\text{eq}}\right),
\end{equation}
where $x \equiv m/T$, the DM yield $Y \equiv \ndm/s$, and the SM entropy density $s(T) = \frac{2 \pi^2}{45}\, \gss\, T^3$, with $\gss(T)$ the number of relativistic degrees of freedom that contribute to the SM entropy.
Taking into account that in a radiation-dominated era $H(T) = \sqrt{\frac{\rR}{3\, M_P^2}} = \frac{\pi}{3} \sqrt{\frac{\gs(T)}{10}}\, \frac{T^2}{M_P}$, with $\gs(T)$ being the number of relativistic degrees of freedom contributing to the SM energy density, Eq.~\eqref{eq:Bzeq2} admits the standard approximate solution
\begin{equation}
    Y_0 \simeq \frac{15}{2 \pi\, \gss} \sqrt{\frac{\gs}{10}}\, \frac{1}{M_P\, \Tfo\, \sv}\,,
\end{equation}
where $Y_0$ corresponds to the DM yield at present, long after freeze-out.
To match the whole observed DM relic density, it is required that
\begin{equation}
    m\, Y_0\, = \Odm h^2 \, \frac{1}{s_0}\,\frac{\rho_c}{h^2} \simeq 4.3 \times 10^{-10}~\text{GeV},
\end{equation}
with $\rho_c \simeq 1.05 \times 10^{-5}\, h^2$~GeV/cm$^3$ being the critical energy density, $s_0\simeq 2.69 \times 10^3$~cm$^{-3}$ the present entropy density~\cite{ParticleDataGroup:2020ssz}, and $\Odm h^2 \simeq 0.12$ the observed DM relic abundance~\cite{Planck:2018vyg}.

The temperature $\Tfo$ at which the DM freeze-out occurs, or equivalently $\xfo \equiv \frac{m}{\Tfo}$, is defined by the equality $\left.\frac{n_\text{eq}\, \sv}{H}\right|_{\xfo} = 1$, and given by
\begin{equation} \label{eq:xfo_RD}
    \xfo = -\frac12\, \mathcal{W}_{-1}\left[- \frac{8 \pi^5}{45} \frac{\gs}{g^2} \frac{1}{\left(M_P\, m\, \sv\right)^2}\right],
\end{equation}
where $\mathcal{W}_{-1}$ is the $-1$ branch of the Lambert function.
The observed DM relic abundance is typically matched for cross-sections at the electroweak scale $\sv = \mathcal{O}\left(10^{-9}\right)$~GeV$^{-2}$ which corresponds to $\mathcal{O}\left(10^{-26}\right)$~cm$^3$/s and $\xfo \sim 25$, with a small logarithmic dependence on the DM mass~\cite{Steigman:2012nb}.

\subsection{Freeze out during Reheating}
Even if the DM freeze-out is typically assumed to occur well {\it after} the end of reheating, when the Universe is radiation dominated, this is not guaranteed.
Alternatively, DM freeze-out could occur {\it during} reheating.
In this case, the SM entropy is not conserved because of the decay of the inflaton into SM particles.
Therefore, instead of Eq.~\eqref{eq:Bzeq2}, it is more convenient to rewrite Eq.~\eqref{eq:Bzeq1} as
\begin{equation} \label{eq:Bzeq3}
    \frac{dN}{da} = -\frac{\sv}{H\,a^4} \left(N^2- N_{\text{eq}}^2\right),
\end{equation}
where $N \equiv a^3\, \ndm$ and $N_\text{eq} \equiv a^3\, n_\text{eq}$. Using Eq.~\eqref{eq:Hubble} for the Hubble parameter during reheating, Eq.~\eqref{eq:Bzeq3} can be analytically solved as
\begin{equation} \label{eq:Narh}
    N(\arh) \simeq \frac{6}{2 + n}\, \frac{\Hrh}{\sv}\, \arh^3 \left(\frac{\arh}{\afo}\right)^\frac{-6}{2 + n},
\end{equation}
where $\afo$ and $\arh$ correspond to the scale factor when $T=\Tfo$ and $T=\Trh$, respectively.
After reheating, namely, for $a > \arh$, the SM entropy is conserved, and therefore $Y_0 \simeq Y(\arh)$, implying
\begin{equation} \label{eq:Y0}
    Y_0 = \frac{N(\arh)}{s(\arh)\, \arh^3} \simeq \frac{6}{2 + n}\, \frac{45}{2\pi^2\, \gss}\, \frac{\Hrh}{\sv\, \Trh^3} \times
    \begin{dcases}
        \left(\frac{\Trh}{\Tfo}\right)^\frac{4}{n - 1} &\text{ for fermionic decay,}\\
        \left(\frac{\Trh}{\Tfo}\right)^4 &\text{ for bosonic decay,}
    \end{dcases}
\end{equation}
for $n < 7$ in the case of fermionic decay.
This assumption will be followed from now on.
We note that, in the era after the freeze-out and before the end of reheating, the DM yield evolves as $Y(T) \propto (a\, T)^{-3}$, which corresponds to
\begin{equation} \label{eq:Yreh}
    Y(T) \propto
    \begin{dcases}
        T^\frac{7 - n}{n - 1} &\text{for fermionic decay,}\\
        T^{1+2\,n} &\text{for bosonic decay.}
    \end{dcases}
\end{equation}

\begin{figure}[t!]
    \def\sepf{0.51}
	\centering
    \includegraphics[scale=\sepf]{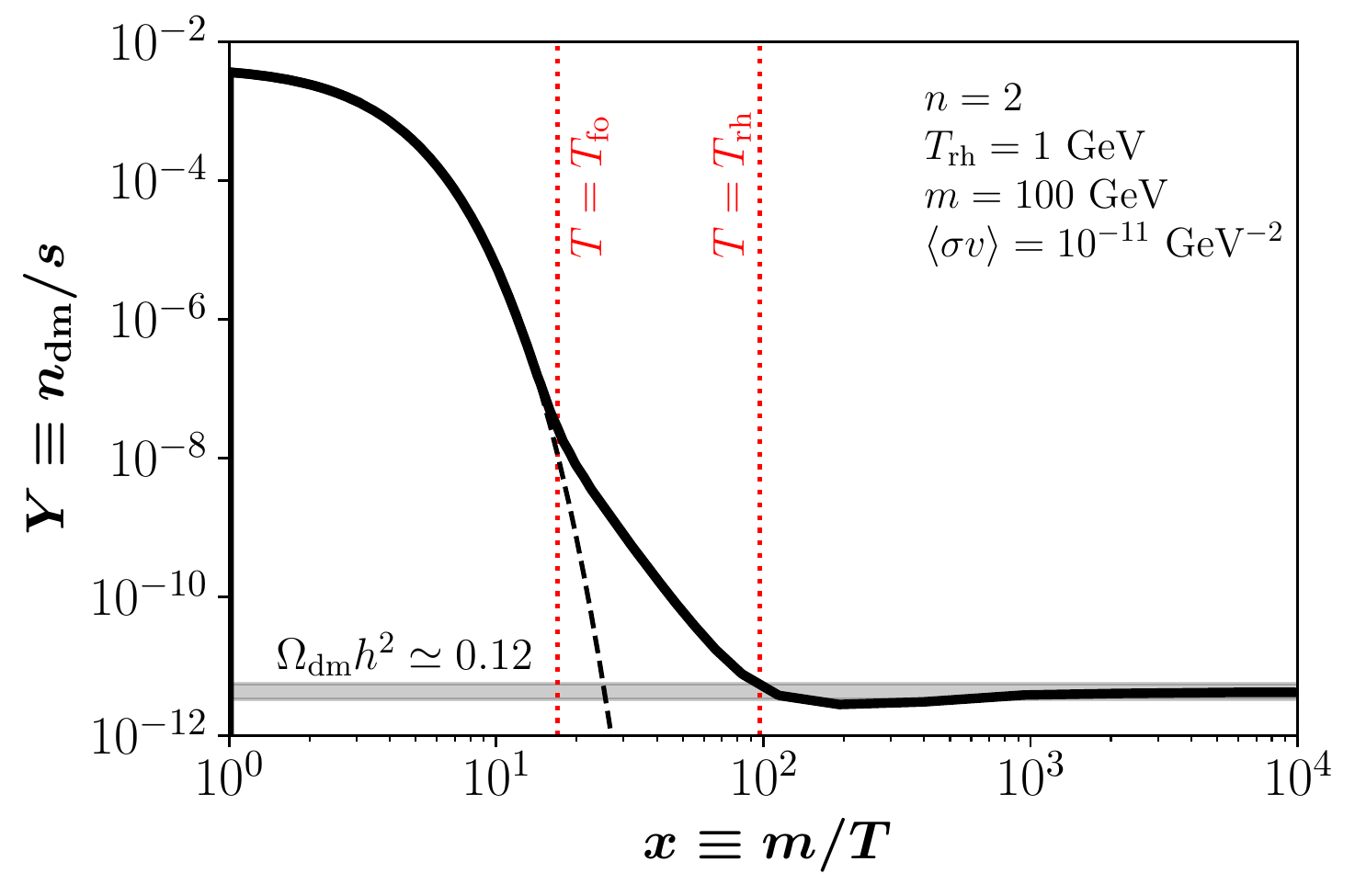}\\
    \includegraphics[scale=\sepf]{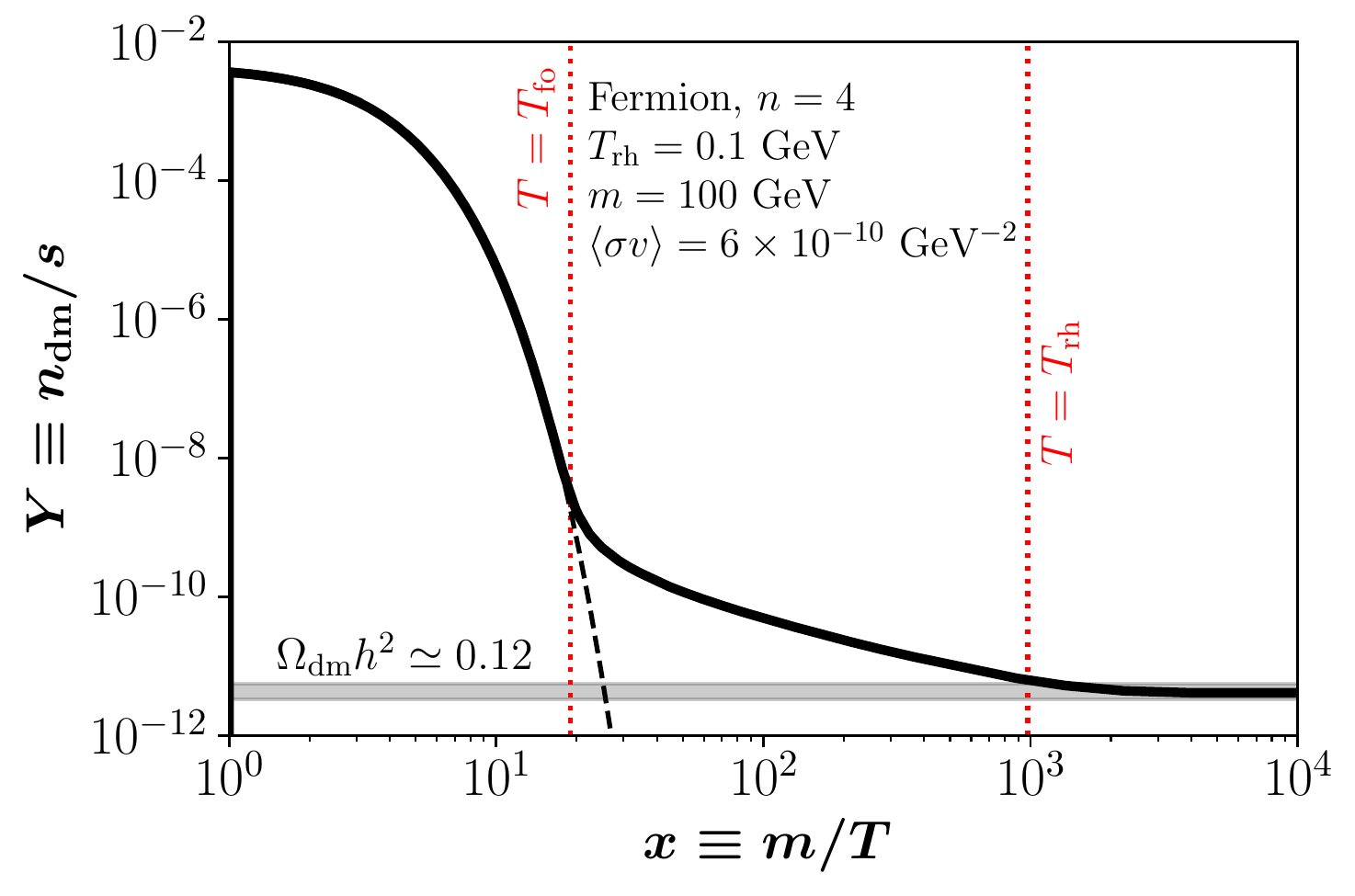}
    \includegraphics[scale=\sepf]{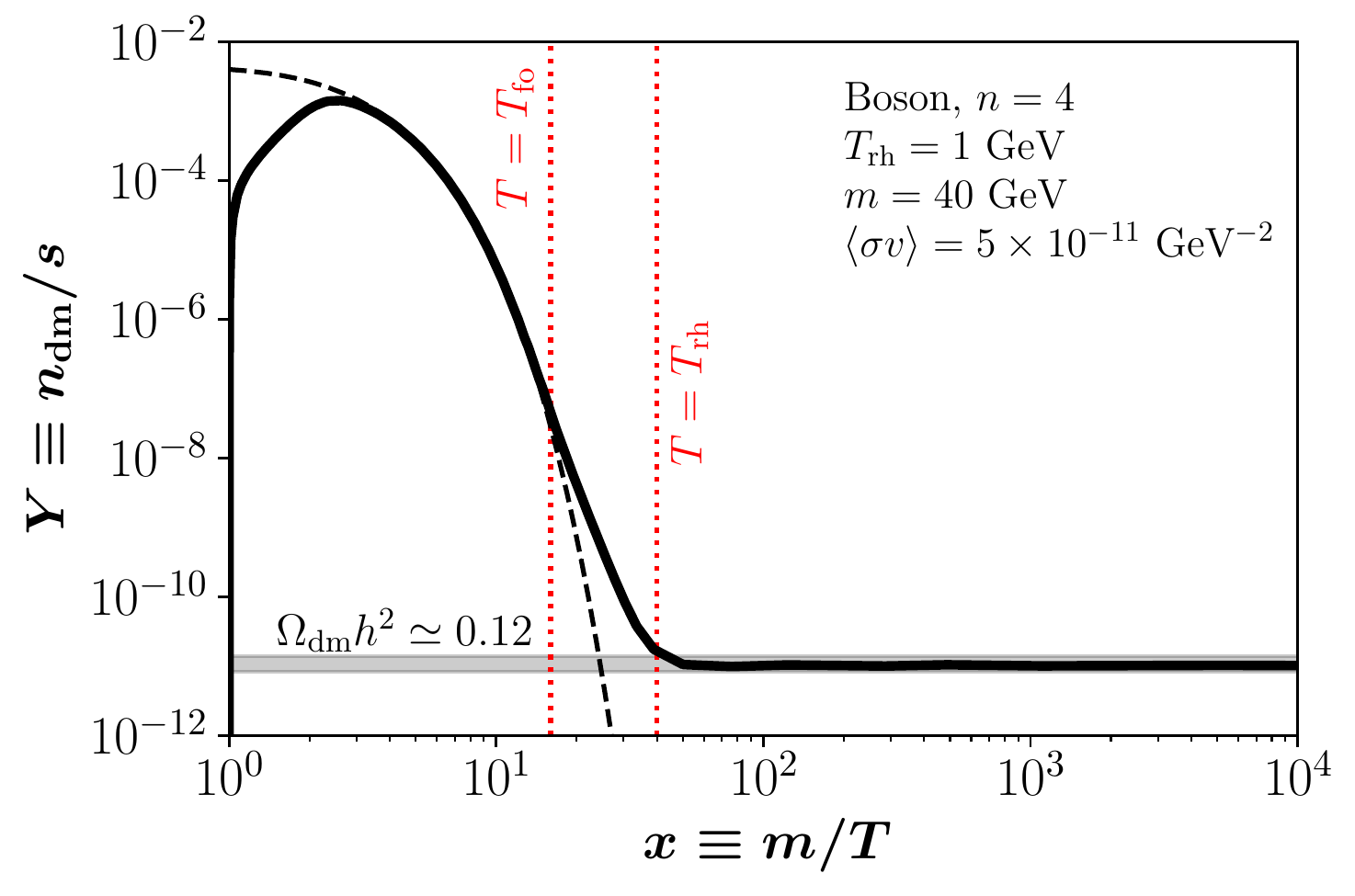}
    \caption{Examples of the evolution of the DM abundance, for different benchmark points described in the text that fit the total observed abundance.
    The solid black lines show the DM yield $Y$, whereas the dashed black lines show the equilibrium density.
    Furthermore, the dotted red vertical lines correspond to $T = \Tfo$ and $T = \Trh$, and the gray horizontal bands depict the observed DM abundance $\Odm h^2\simeq 0.12$.}
	\label{fig:FO}
\end{figure} 
Examples of the evolution of DM abundance are shown in Fig.~\ref{fig:FO}, for reference points that fit the total observed abundance: the upper panel corresponds to $n = 2$, $\Trh = 1$~GeV, $m = 100$~GeV and $\sv = 10^{-11}$~GeV$^{-2}$, the lower left panel corresponds to a fermionic decay with $n = 4$, $\Trh = 0.1$~GeV, $m = 100$~GeV and $\sv = 6 \times 10^{-10}$~GeV$^{-2}$, and the lower right panel corresponds to a bosonic decay with $n = 4$, $\Trh = 1$~GeV, $m = 40$~GeV and $\sv = 5 \times 10^{-11}$~GeV$^{-2}$.
The solid black lines depict the evolution of the DM yield $Y$, while the dashed black lines show the equilibrium yield $Y_{\text{eq}}$.
Furthermore, the red dotted vertical lines show $T = \Tfo$ and $T = \Trh$, and the gray horizontal bands correspond to the observed DM abundance $\Odm h^2\simeq 0.12$~\cite{Planck:2018vyg}. 

The evolution of the DM yield depicted in Fig.~\ref{fig:FO} is obtained by numerically solving the Boltzmann equations~\eqref{eq:drhodt}, \eqref{eq:rR}, \eqref{eq:Bzeq1} and the Friedmann equation~\eqref{eq:Htotal}.
We have checked that the analytical estimate of the asymptotic value of $Y_0$ shown in Eq.~\eqref{eq:Y0} agrees well with the numerical result.
Furthermore, the evolution of the DM yield for $\Tfo \gg T \gg \Trh$ (that is, $Y \propto x^{-5}$ for $n = 2$, and $Y \propto x^{-1}$ or $Y \propto x^{-9}$ for fermionic or bosonic decays, respectively, and $n = 4$) also matches the analytical expressions of Eq.~\eqref{eq:Yreh}.
It is worth noticing that in the lower right panel, the DM only reaches chemical equilibrium with the SM bath at $x \simeq 3$.

\begin{figure}[t!]
    \def\sepf{0.51}
	\centering
    \includegraphics[scale=\sepf]{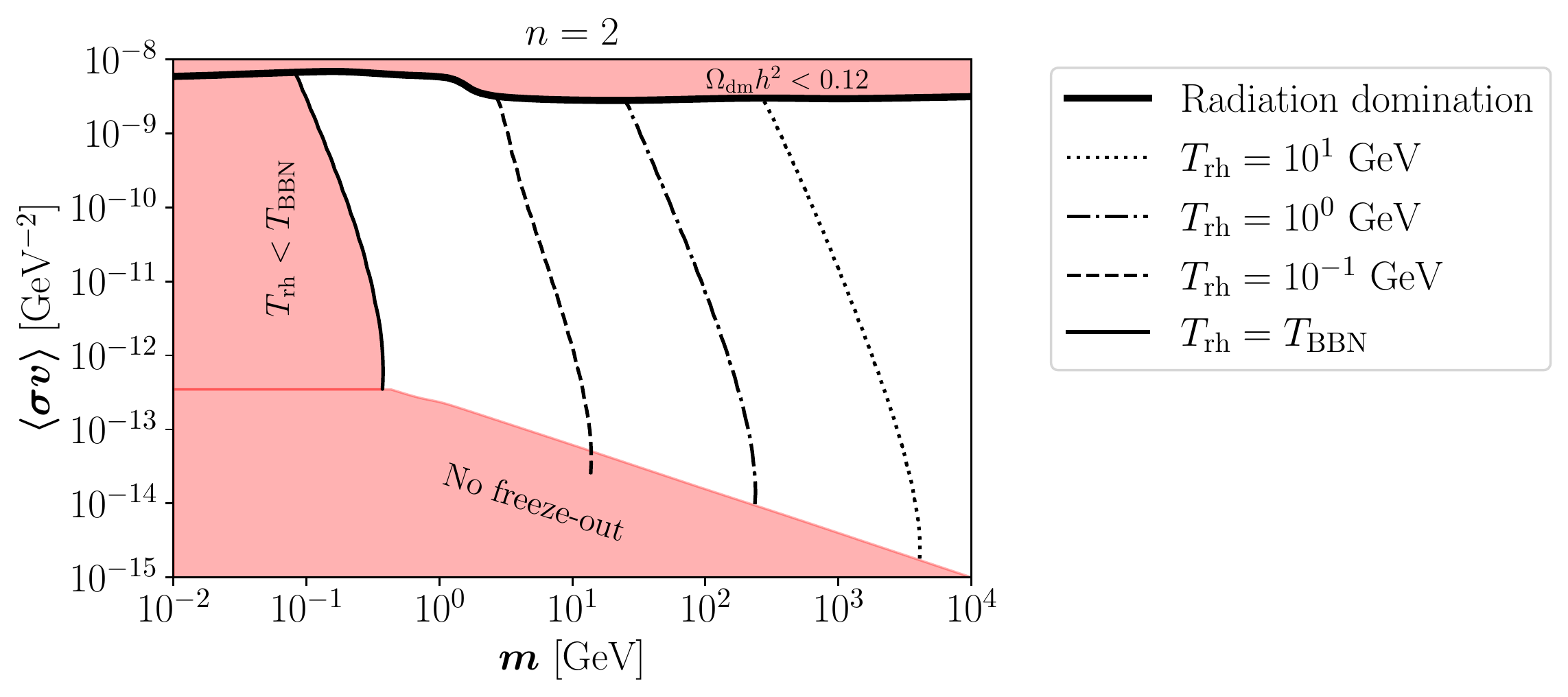}
    \includegraphics[scale=\sepf]{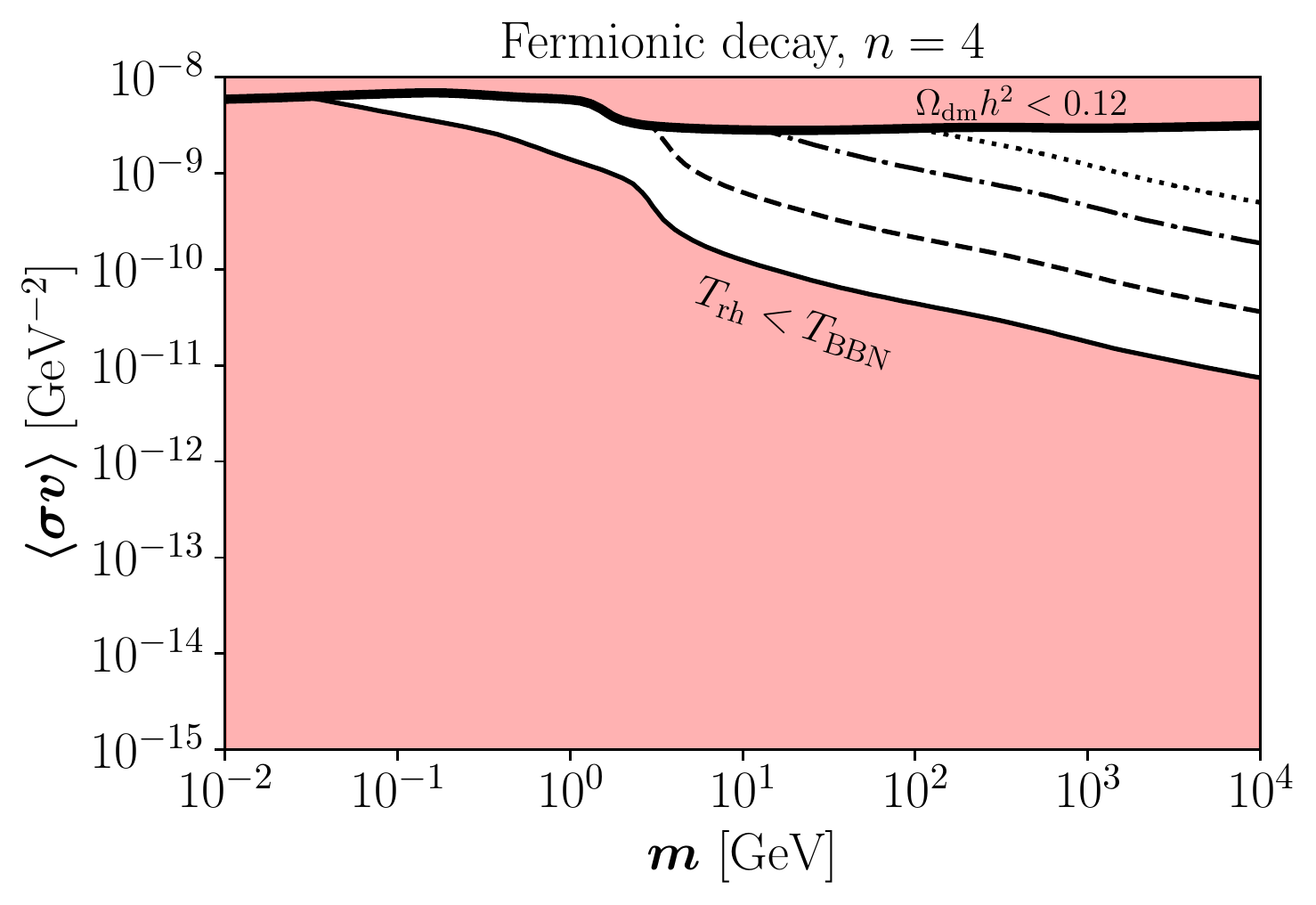}
    \includegraphics[scale=\sepf]{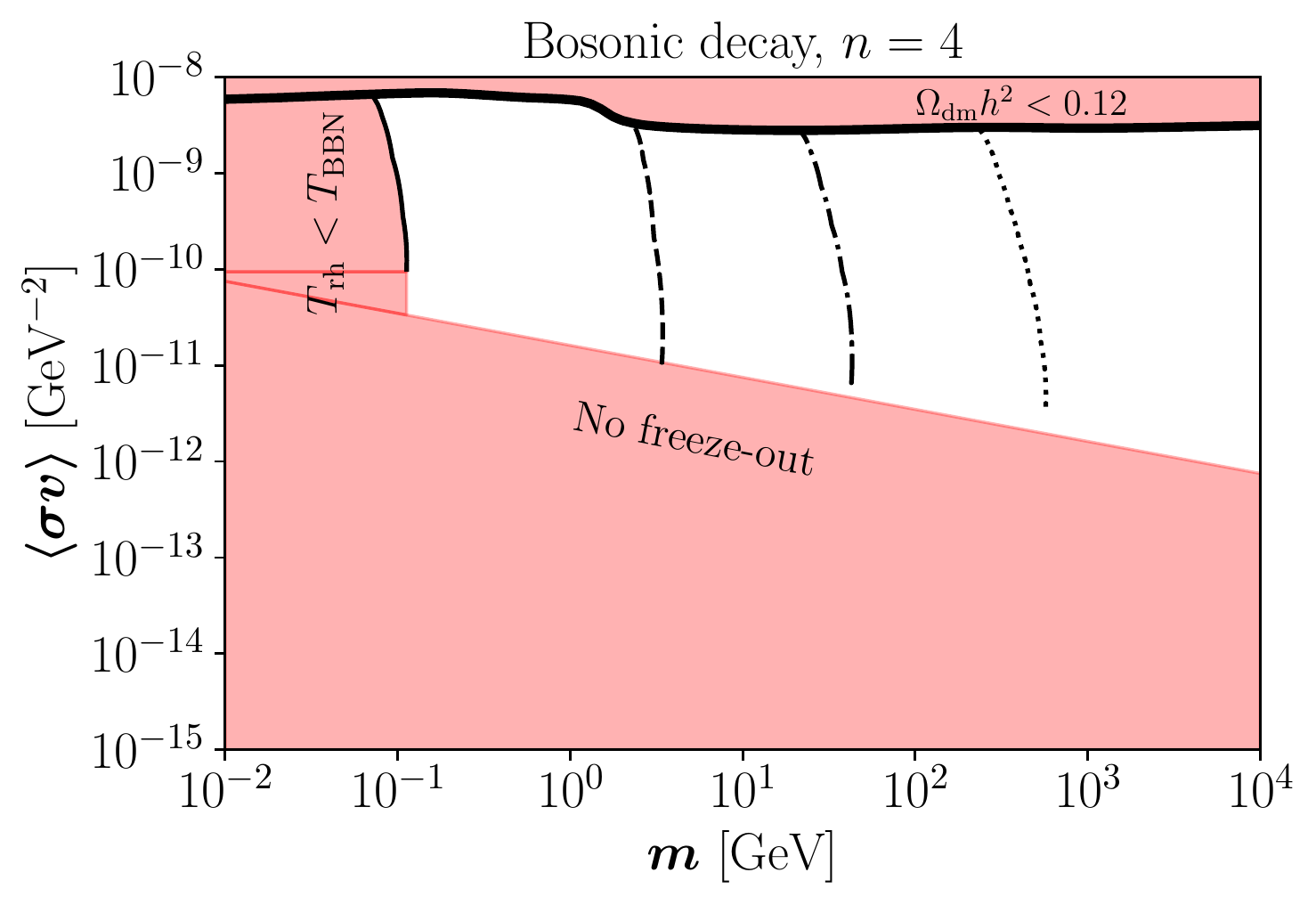}
    \caption{Parameter space that matches the whole observed DM abundance, for $\Trh = T_\text{BBN}$ (solid lines), $10^{-1}$~GeV (dashed), $10^0$~GeV (dash-dotted), and $10^1$~GeV (dotted).
    The upper panel corresponds to $n = 2$, while the lower to $n = 4$ and fermionic (left) or bosonic (right) decays of the inflaton.
    The thick black lines correspond to the standard scenario, where DM freeze-out in the radiation-dominated era.
    The red areas are disregarded as they generate a DM underabundance ($\Odm < 0.12$), the reheating occurs after BBN ($\Trh < T_\text{BBN}$), or DM does not reach chemical equilibrium with the SM (`No freeze-out').
    }
	\label{fig:m-sv}
\end{figure} 
The contours for the reheating temperature required to match the whole measured DM abundance are shown in Fig.~\ref{fig:m-sv}, in the parameter space $[m,\, \sv]$, for bosonic and fermionic decays of the inflaton, and different values of $n$.
The thick black lines correspond to the standard case, where DM freezes out in a radiation-dominated era, whereas the thin black lines to $\Trh = T_\text{BBN}$, $10^{-1}$~GeV, $10^0$~GeV and $10^1$~GeV.
Some constraints apply and are shown as red regions: $i)$ Above the thick black line, corresponding to higher cross sections, DM decouples very late and is therefore underproduced ($\Odm h^2 < 0.12$). $ii)$ Reheating temperatures below $T_\text{BBN}$ are in conflict with cosmological observations, which corresponds to $m \lesssim \mathcal{O}(10^{-1})$~GeV. And $iii)$, very small thermally averaged cross sections $\sv$ could not be enough to guarantee chemical equilibrium between the dark and visible sectors.
It is worth mentioning that in this case, labeled `No freeze-out', the whole DM relic abundance could be matched, however, the production would not correspond to the WIMP mechanism, but rather to a FIMP scenario, with the usual strong dependence on initial conditions.
This case will not be considered in the present analysis.

In Fig.~\ref{fig:m-sv}, it can be seen that a DM freeze-out during reheating allows exploring smaller cross-sections $\sv$ compared to the usual case where the freeze-out occurs in the standard radiation-dominated case. For example, with $n = 2$ and $\Trh = 10$~GeV, $\sv$ can be as small as $\mathcal{O}(10^{-15})$~GeV$^{-2}$ for a DM mass of $\sim 7$~TeV. 
In a scenario with larger $\Trh$, even smaller $\sv$ are allowed for heavier WIMPs.
We also note that for a fixed $\Trh$, $\sv$ decreases with increasing $m$. This is because for heavier WIMPs, freeze-out occurs earlier (i.e., at a larger $\Tfo$), which implies a smaller $\sv$. Finally, it should be noted that with increasing $n$, the parameter space shrinks, as depicted in Fig.~\ref{fig:m-sv}. The main reason is that the dilution effect becomes less prominent with larger $n$ (that is, the factor $\left(\arh/\afo\right)^\frac{-6}{2 + n}$ in Eq.~\eqref{eq:Narh}), and hence a larger $\sv$ is needed to compensate for this. A larger $\sv$ corresponds to a later freeze-out, where the WIMP densities are smaller, and therefore a smaller dilution is required.

\begin{figure}[t!]
    \def\sepf{0.51}
	\centering
    \includegraphics[scale=\sepf]{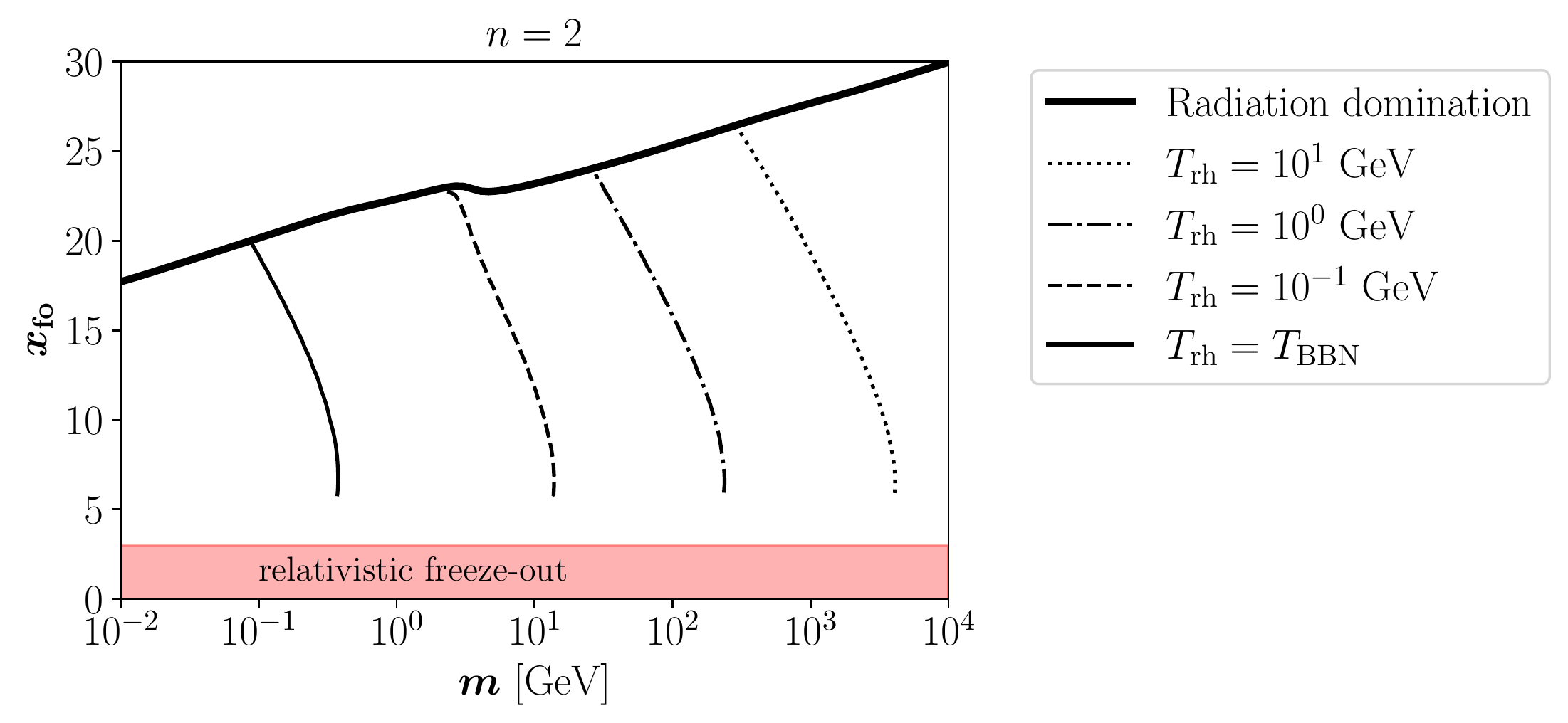}
    \includegraphics[scale=\sepf]{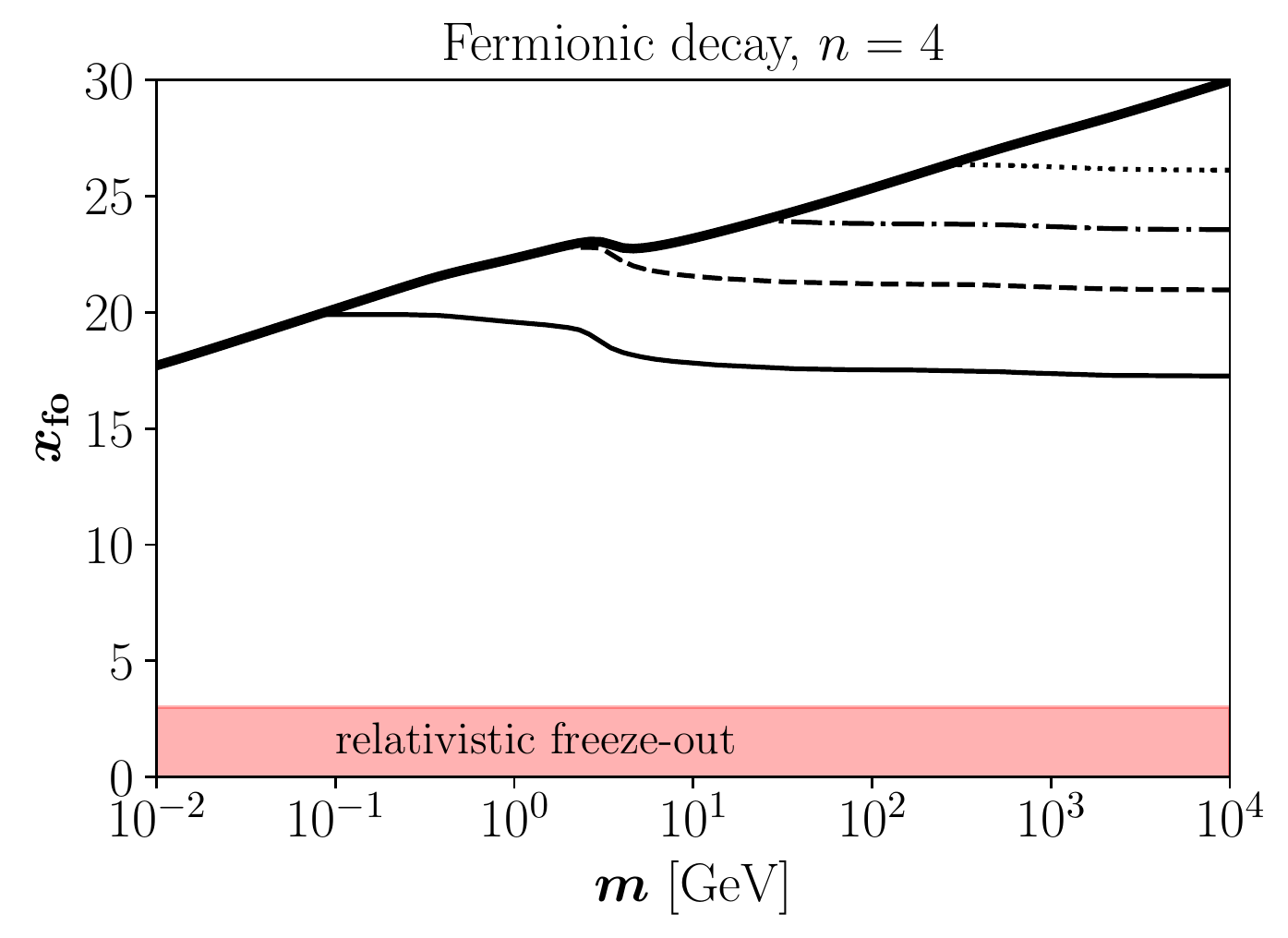}
    \includegraphics[scale=\sepf]{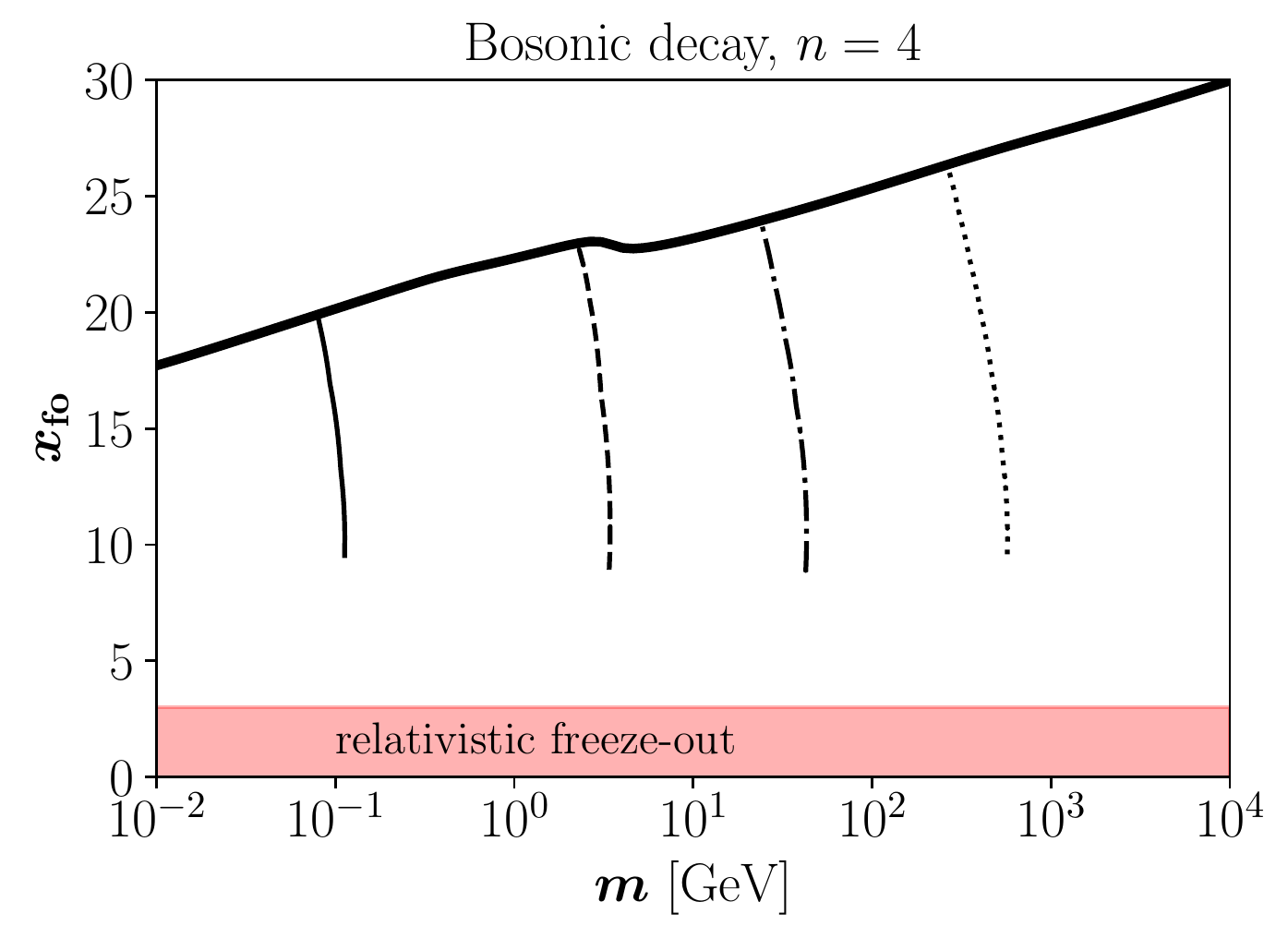}
    \caption{Freeze-out temperature required to match the whole observed DM abundance, as a function of the DM mass, for $\Trh = T_\text{BBN}$ (solid lines), $10^{-1}$~GeV (dashed lines), $10^0$~GeV (dotted lines in dashes) and $10^1$~GeV (dotted lines).
    The upper panel corresponds to $n = 2$, while the lower left to a fermionic decay with $n = 4$, and the lower right to a bosonic decay with $n = 4$.
    The thick black lines represent the standard freeze-out in radiation domination.
    }
	\label{fig:xfo}
\end{figure} 
In the present case, DM freezes out during reheating, at a temperature given by
\begin{equation}
    \xfo \simeq
    \begin{dcases}
        \frac{3 + n}{2\, (1 - n)}\, \mathcal{W}_{-1}\left[\frac{2\, (1 - n)}{3 + n} \left(\frac{g^2}{(2 \pi)^3}\, \frac{m^2\, \Trh^4\, \sv^2}{\Hrh^2}\right)^\frac{1 - n}{3 + n} \left(\frac{m}{\Trh}\right)^\frac{4}{3 + n}\right] &\text{for fermions,}\\
        \frac{3 - 4\, n}{2}\, \mathcal{W}_{-1}\left[\frac{2}{3 - 4\, n} \left(\frac{g^2}{(2 \pi)^3}\, \frac{m^2\, \Trh^4 \sv^2}{\Hrh^2}\right)^\frac{1}{3 - 4 n} \left(\frac{m}{\Trh}\right)^\frac{4\, (1 - n)}{3 - 4 n}\right] &\text{for bosons.}
    \end{dcases}
\end{equation}
Figure~\ref{fig:xfo} shows the freeze-out temperature required to match the entire observed DM abundance, as a function of the DM mass, for $\Trh = T_\text{BBN}$ (solid lines), $10^{-1}$~GeV (dashed lines), $10^0$~GeV (dash-dotted lines) and $10^1$~GeV (dotted lines).
The upper panel corresponds to $n = 2$, the lower left to a fermionic decay with $n = 4$, and the lower right to a bosonic decay with $n = 4$.
The thick black lines represent the standard freeze-out in radiation domination; cf. Eq.~\eqref{eq:xfo_RD}.
Additionally, the lower red bands depict a relativistic freeze-out, i.e. $\xfo \leq 3$.
Note that for $n = 2$, the freeze-out temperature can be as low as $\xfo \sim 6$ due to the entropy dilution effect. 
For $n = 2$ and $n = 4$ (bosonic), the lines are cut to ensure that WIMPs thermalize; lower values of $\xfo$ would correspond to the `No freeze-out' regime depicted in Fig.~\ref{fig:m-sv}. For a fermionic decay with $n = 4$, $\xfo$ tends to be independent of $m$, with only a small dependence on the relativistic degrees of freedom. 
Note that the temperature during reheating (for the fermionic case) tends to feature the same scaling as the free radiation for large values of $n$, cf. Eq.~\eqref{eq:Tfer}; this implies that the corresponding freeze-out behavior becomes similar to the standard case. Actually, such a tendency can also be seen in Fig.~\ref{fig:xfo}.
However, for the bosonic case, due to a different temperature scaling (cf. Eq.~\eqref{eq:TBos}), the allowed parameter space becomes distinct from the fermionic case with an increase of $n$. Note that, as argued earlier, in both fermonic and bosonic cases, with increasing $n$, a larger $\sv$ (corresponding to a later freeze-out with lower $\Tfo$) is needed, implying that $\xfo$ increases with $n$, as depicted in Fig.~\ref{fig:xfo}.

\begin{figure}[t!]
    \def\sepf{0.47}
	\centering
    \includegraphics[scale=\sepf]{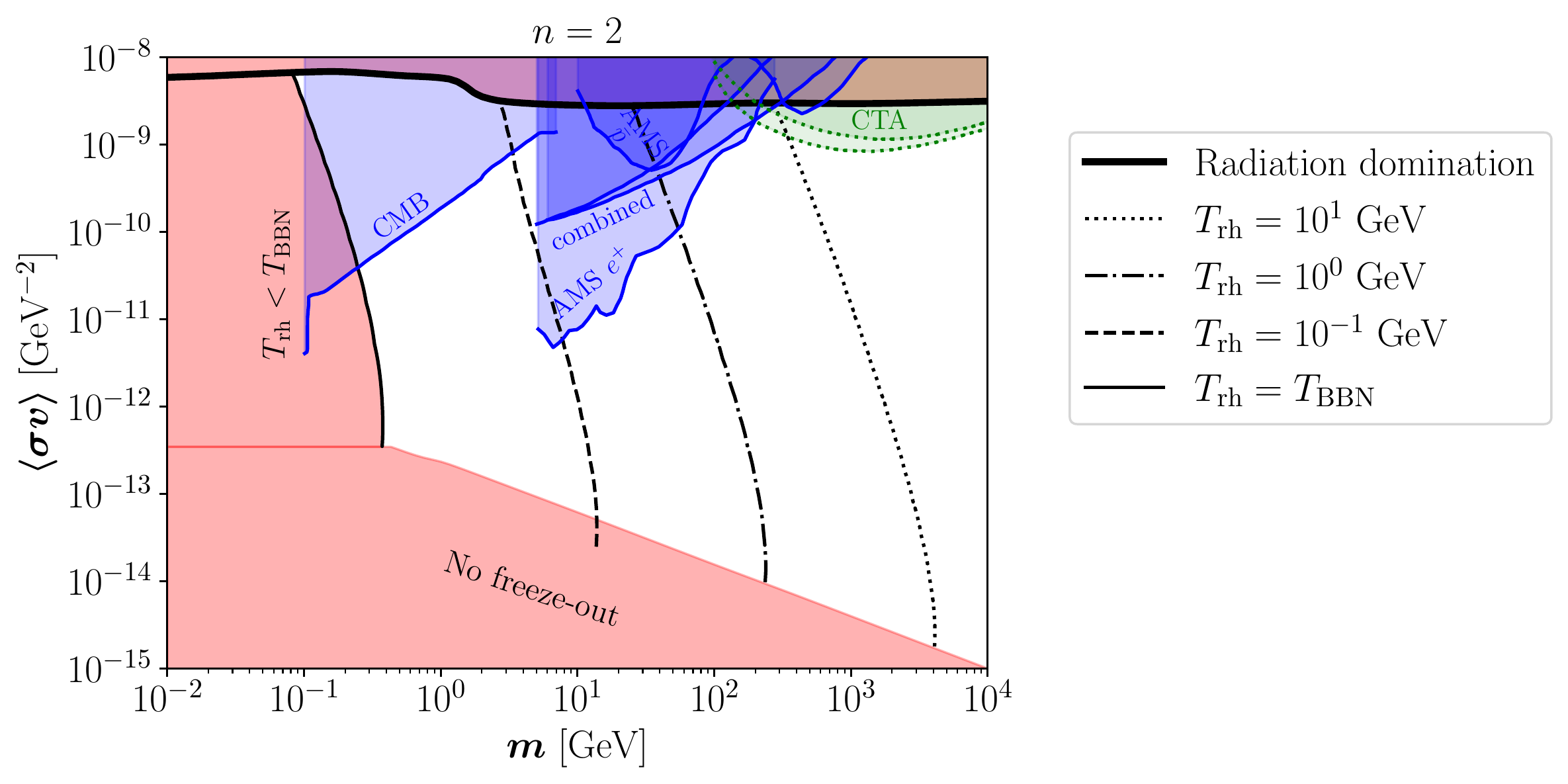}
    \includegraphics[scale=\sepf]{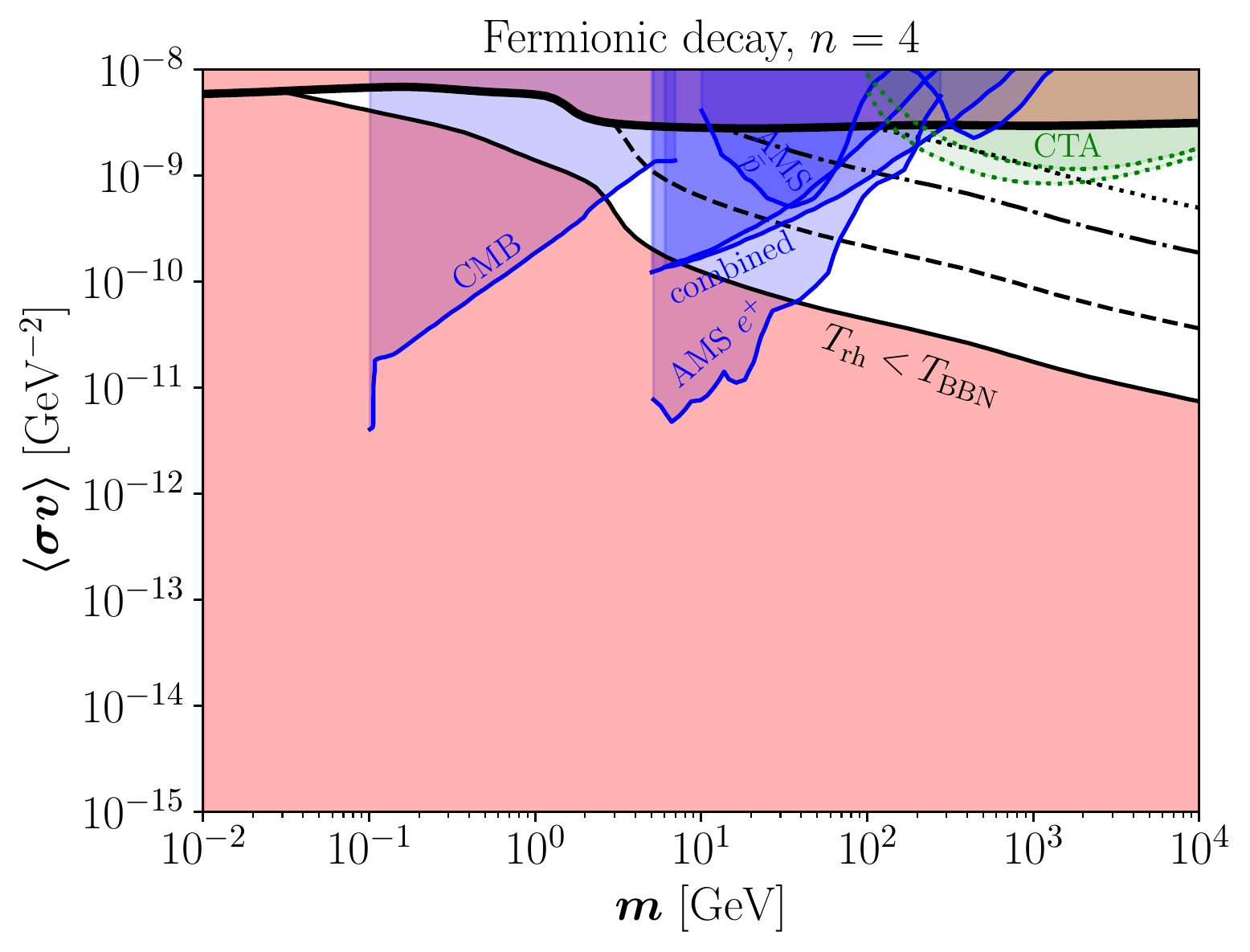}
    \includegraphics[scale=\sepf]{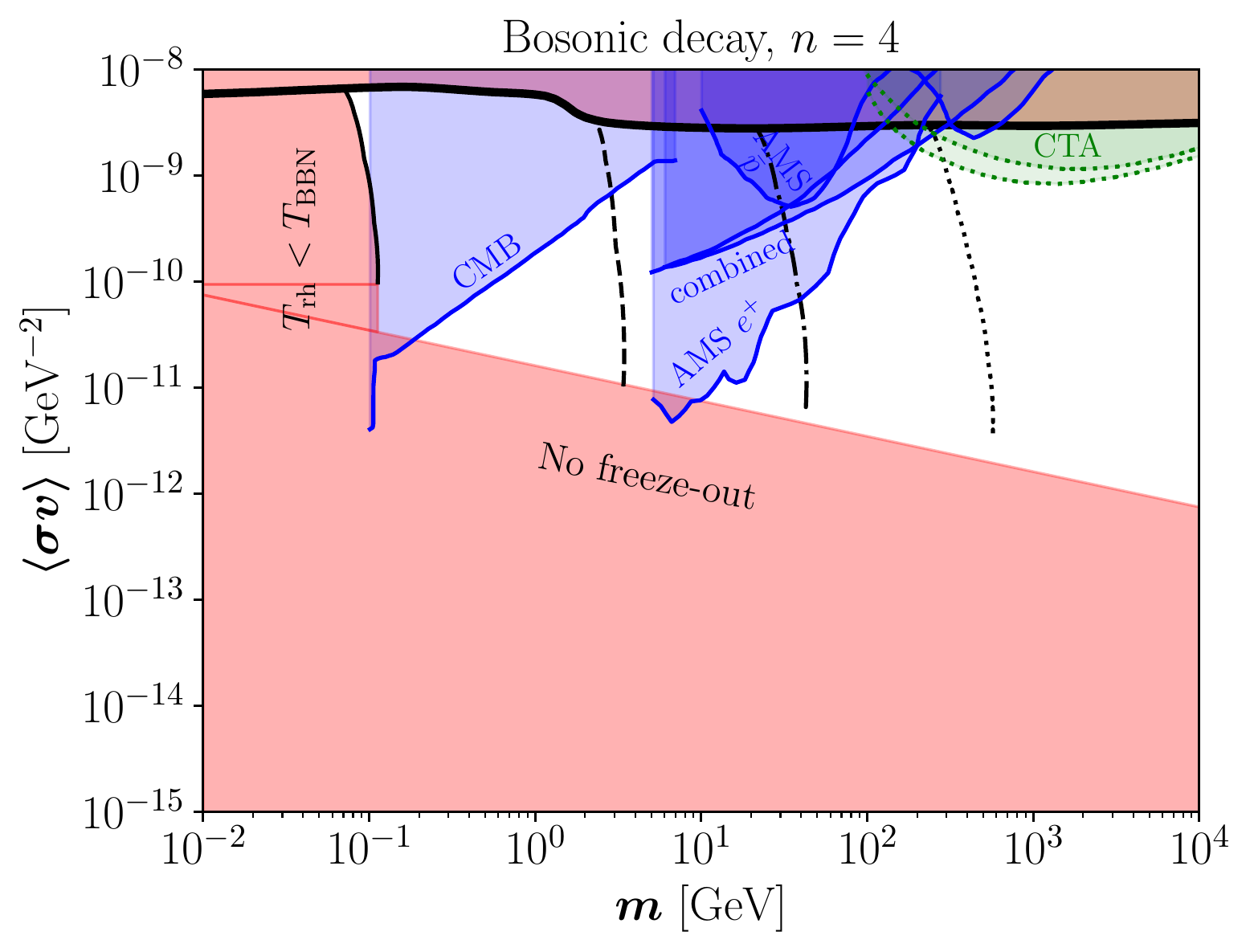}
    \caption{Same as Fig.~\ref{fig:m-sv} with current experimental bounds (blue) and future sensitivities (green).}
	\label{fig:sv-bound}
\end{figure} 
Before closing this section, we comment on some present and future possibilities of testing the current scenario using the different channels offered by DM indirect detection experiments.
In addition to the information given in Fig.~\ref{fig:m-sv}, Fig.~\ref{fig:sv-bound} also shows in blue present constraints coming from: $i)$ CMB spectral distortions from WIMP annihilation into charged particles~\cite{Leane:2018kjk} (labeled `CMB'), $ii)$ the combined analysis (labeled `combined') of $\gamma$-ray data using Fermi-LAT, HAWC, H.E.S.S., MAGIC, and VERITAS by considering WIMP annihilation to $\tau^+ \tau^-$ (upper line) and $b \bar b$ (lower line)~\cite{Hess:2021cdp}, and $iii)$ AMS-02 measurements of antimatter in cosmic rays, in the positron~\cite{Bergstrom:2013jra} and the antiproton~\cite{Calore:2022stf} channels.\footnote{It is interesting to recall that the bounds presented depends on the DM annihilation channels, and suffer from large uncertainties arising from e.g. assumptions on the DM density profile, the local DM density, and the propagation of charged particles in the interstellar medium~\cite{Zemp:2008gw, Pato:2010yq, Bernal:2014mmt, Boudaud:2014dta, Bernal:2015oyn, Bernal:2016guq, Benito:2016kyp}.}
Additionally, Fig.~\ref{fig:sv-bound} shows in green the projected sensitivity of the ground-based CTA experiment, which could be capable of testing WIMP DM at the TeV scale and above. The upper (lower) green dotted line shows its expected sensitivity for WIMP annihilation to $b \bar{b}$ ($W^+ W^-$)~\cite{CTA:2020qlo}. Note that CTA might be able to probe a large fraction of the TeV-scale WIMPs in the standard scenario, and even a chunk of the parameter space where the freeze-out occurs during reheating, for both fermionic and bosonic decays.

\section{Conclusions} \label{sec:concl}
Despite the huge experimental efforts over the past decades, the nature of dark matter (DM) remains one of the most challenging and fundamental questions in particle physics. In particular, scenarios where DM is a weakly interacting massive particle (WIMP) reaching thermal equilibrium with the standard model (SM) bath have received immense attention both theoretically and experimentally, but unfortunately no overwhelming evidence for WIMP DM has been found. This motivates a quest beyond the standard WIMP paradigm. An alternative is to renounce the thermal nature of DM, as in the FIMP paradigm. However, these kinds of scenario suffer from a strong dependence on the largely unknown initial conditions after cosmological inflation.

In this work, we focus on another avenue where DM remains a thermal relic, but its freeze-out (that is, its departure from chemical equilibrium) occurs {\it during} (and not after) inflationary reheating. In particular, we investigated a scenario where the inflaton field $\phi$ coherently oscillates at the bottom of a generic potential $V(\phi) \propto \phi^n$, while decaying into SM particles to reheat the Universe.
Depending on the details of the reheating (i.e. the shape of the potential and the coupling between the inflaton and the SM particles), the inflaton and the SM energy density could feature a distinct evolution. Consequently, in this case the behavior of WIMP freeze-out differs from the case where it occurs well after reheating. Due to the dilution effect during reheating, smaller thermally averaged cross-sections $\sv$ becomes compatible with the measured DM relic abundance.

In particular, if DM freezes-out after reheating, one requires $\sv \sim \mathcal{O}(10^{-9})$~GeV$^{-2} \sim \mathcal{O}(10^{-26})$~cm$^3$/s to match the observed DM abundance, and a DM freeze-out temperature $\xfo \equiv m/\Tfo \sim 25$.
Instead, if the freeze-out occurs during reheating while the inflaton oscillates in a quadratic potential ($n = 2$), if $\Trh \sim 1$~GeV, $\sv$ could be of the order $\mathcal{O}(10^{-15})$~GeV$^{-2}$ for a WIMP at the TeV scale.
Interestingly, a sizable part of the parameter space is in the range of current experiments and future proposals.
For instance, for TeV-scale WIMPs, the upcoming CTA could probe the expected cross-sections for DM decoupling after reheating, and a portion of the plane $[m,\, \sv]$ relevant for the case where DM decouples during reheating.

\section*{Acknowledgments}
We would like to thank Manuel Drees once again for enlightening discussions. NB received funding from the Patrimonio Autónomo - Fondo Nacional de Financiamiento para la Ciencia, la Tecnología y la Innovación Francisco José de Caldas (MinCiencias - Colombia) grants 80740-465-2020 and 80740-492-2021.
NB is also funded by the Spanish FEDER/MCIU-AEI under grant FPA2017-84543-P.
This project has received funding/support from the European Union's Horizon 2020 research and innovation program under the Marie Skłodowska-Curie grant agreement No 860881-HIDDeN. 

\bibliographystyle{JHEP}
\bibliography{biblio}

\providecommand{\href}[2]{#2}\begingroup\raggedright\begin{thebibliography}{100}

\bibitem{Bertone:2016nfn}
G.~Bertone and D.~Hooper, \emph{{History of dark matter}},
  \href{https://doi.org/10.1103/RevModPhys.90.045002}{\emph{Rev. Mod. Phys.}
  {\bfseries 90} (2018) 045002}
  [\href{https://arxiv.org/abs/1605.04909}{{\ttfamily 1605.04909}}].

\bibitem{Planck:2018vyg}
{\scshape Planck} collaboration, \emph{{Planck 2018 results. VI. Cosmological
  parameters}},
  \href{https://doi.org/10.1051/0004-6361/201833910}{\emph{Astron. Astrophys.}
  {\bfseries 641} (2020) A6}
  [\href{https://arxiv.org/abs/1807.06209}{{\ttfamily 1807.06209}}].

\bibitem{Drees:2018hzm}
M.~Drees, \emph{{Dark Matter Theory}},
  \href{https://doi.org/10.22323/1.340.0730}{\emph{PoS} {\bfseries ICHEP2018}
  (2019) 730} [\href{https://arxiv.org/abs/1811.06406}{{\ttfamily
  1811.06406}}].

\bibitem{Arcadi:2017kky}
G.~Arcadi, M.~Dutra, P.~Ghosh, M.~Lindner, Y.~Mambrini, M.~Pierre et~al.,
  \emph{{The waning of the WIMP? A review of models, searches, and
  constraints}},
  \href{https://doi.org/10.1140/epjc/s10052-018-5662-y}{\emph{Eur. Phys. J. C}
  {\bfseries 78} (2018) 203}
  [\href{https://arxiv.org/abs/1703.07364}{{\ttfamily 1703.07364}}].

\bibitem{Roszkowski:2017nbc}
L.~Roszkowski, E.M.~Sessolo and S.~Trojanowski, \emph{{WIMP dark matter
  candidates and searches\textemdash{}current status and future prospects}},
  \href{https://doi.org/10.1088/1361-6633/aab913}{\emph{Rept. Prog. Phys.}
  {\bfseries 81} (2018) 066201}
  [\href{https://arxiv.org/abs/1707.06277}{{\ttfamily 1707.06277}}].

\bibitem{Steigman:2012nb}
G.~Steigman, B.~Dasgupta and J.F.~Beacom, \emph{{Precise Relic WIMP Abundance
  and its Impact on Searches for Dark Matter Annihilation}},
  \href{https://doi.org/10.1103/PhysRevD.86.023506}{\emph{Phys. Rev. D}
  {\bfseries 86} (2012) 023506}
  [\href{https://arxiv.org/abs/1204.3622}{{\ttfamily 1204.3622}}].

\bibitem{McDonald:2001vt}
J.~McDonald, \emph{{Thermally generated gauge singlet scalars as
  selfinteracting dark matter}},
  \href{https://doi.org/10.1103/PhysRevLett.88.091304}{\emph{Phys. Rev. Lett.}
  {\bfseries 88} (2002) 091304}
  [\href{https://arxiv.org/abs/hep-ph/0106249}{{\ttfamily hep-ph/0106249}}].

\bibitem{Choi:2005vq}
K.-Y.~Choi and L.~Roszkowski, \emph{{E-WIMPs}},
  \href{https://doi.org/10.1063/1.2149672}{\emph{AIP Conf. Proc.} {\bfseries
  805} (2005) 30} [\href{https://arxiv.org/abs/hep-ph/0511003}{{\ttfamily
  hep-ph/0511003}}].

\bibitem{Kusenko:2006rh}
A.~Kusenko, \emph{{Sterile neutrinos, dark matter, and the pulsar velocities in
  models with a Higgs singlet}},
  \href{https://doi.org/10.1103/PhysRevLett.97.241301}{\emph{Phys. Rev. Lett.}
  {\bfseries 97} (2006) 241301}
  [\href{https://arxiv.org/abs/hep-ph/0609081}{{\ttfamily hep-ph/0609081}}].

\bibitem{Petraki:2007gq}
K.~Petraki and A.~Kusenko, \emph{{Dark-matter sterile neutrinos in models with
  a gauge singlet in the Higgs sector}},
  \href{https://doi.org/10.1103/PhysRevD.77.065014}{\emph{Phys. Rev. D}
  {\bfseries 77} (2008) 065014}
  [\href{https://arxiv.org/abs/0711.4646}{{\ttfamily 0711.4646}}].

\bibitem{Hall:2009bx}
L.J.~Hall, K.~Jedamzik, J.~March-Russell and S.M.~West, \emph{{Freeze-In
  Production of FIMP Dark Matter}},
  \href{https://doi.org/10.1007/JHEP03(2010)080}{\emph{JHEP} {\bfseries 03}
  (2010) 080} [\href{https://arxiv.org/abs/0911.1120}{{\ttfamily 0911.1120}}].

\bibitem{Elahi:2014fsa}
F.~Elahi, C.~Kolda and J.~Unwin, \emph{{UltraViolet Freeze-in}},
  \href{https://doi.org/10.1007/JHEP03(2015)048}{\emph{JHEP} {\bfseries 03}
  (2015) 048} [\href{https://arxiv.org/abs/1410.6157}{{\ttfamily 1410.6157}}].

\bibitem{Bernal:2017kxu}
N.~Bernal, M.~Heikinheimo, T.~Tenkanen, K.~Tuominen and V.~Vaskonen, \emph{{The
  Dawn of FIMP Dark Matter: A Review of Models and Constraints}},
  \href{https://doi.org/10.1142/S0217751X1730023X}{\emph{Int. J. Mod. Phys. A}
  {\bfseries 32} (2017) 1730023}
  [\href{https://arxiv.org/abs/1706.07442}{{\ttfamily 1706.07442}}].

\bibitem{Allahverdi:2020bys}
R.~Allahverdi et~al., \emph{{The First Three Seconds: a Review of Possible
  Expansion Histories of the Early Universe}},
  \href{https://doi.org/10.21105/astro.2006.16182}{\emph{Open J.Astrophys.}
  {\bfseries 04} (2021) } [\href{https://arxiv.org/abs/2006.16182}{{\ttfamily
  2006.16182}}].

\bibitem{Arias:2019uol}
P.~Arias, N.~Bernal, A.~Herrera and C.~Maldonado, \emph{{Reconstructing
  Non-standard Cosmologies with Dark Matter}},
  \href{https://doi.org/10.1088/1475-7516/2019/10/047}{\emph{JCAP} {\bfseries
  10} (2019) 047} [\href{https://arxiv.org/abs/1906.04183}{{\ttfamily
  1906.04183}}].

\bibitem{Kane:2015jia}
G.~Kane, K.~Sinha and S.~Watson, \emph{{Cosmological Moduli and the
  Post-Inflationary Universe: A Critical Review}},
  \href{https://doi.org/10.1142/S0218271815300220}{\emph{Int. J. Mod. Phys. D}
  {\bfseries 24} (2015) 1530022}
  [\href{https://arxiv.org/abs/1502.07746}{{\ttfamily 1502.07746}}].

\bibitem{Co:2015pka}
R.T.~Co, F.~D'Eramo, L.J.~Hall and D.~Pappadopulo, \emph{{Freeze-In Dark Matter
  with Displaced Signatures at Colliders}},
  \href{https://doi.org/10.1088/1475-7516/2015/12/024}{\emph{JCAP} {\bfseries
  12} (2015) 024} [\href{https://arxiv.org/abs/1506.07532}{{\ttfamily
  1506.07532}}].

\bibitem{Davoudiasl:2015vba}
H.~Davoudiasl, D.~Hooper and S.D.~McDermott, \emph{{Inflatable Dark Matter}},
  \href{https://doi.org/10.1103/PhysRevLett.116.031303}{\emph{Phys. Rev. Lett.}
  {\bfseries 116} (2016) 031303}
  [\href{https://arxiv.org/abs/1507.08660}{{\ttfamily 1507.08660}}].

\bibitem{Randall:2015xza}
L.~Randall, J.~Scholtz and J.~Unwin, \emph{{Flooded Dark Matter and S Level
  Rise}}, \href{https://doi.org/10.1007/JHEP03(2016)011}{\emph{JHEP} {\bfseries
  03} (2016) 011} [\href{https://arxiv.org/abs/1509.08477}{{\ttfamily
  1509.08477}}].

\bibitem{Berlin:2016vnh}
A.~Berlin, D.~Hooper and G.~Krnjaic, \emph{{PeV-Scale Dark Matter as a Thermal
  Relic of a Decoupled Sector}},
  \href{https://doi.org/10.1016/j.physletb.2016.06.037}{\emph{Phys. Lett. B}
  {\bfseries 760} (2016) 106}
  [\href{https://arxiv.org/abs/1602.08490}{{\ttfamily 1602.08490}}].

\bibitem{Tenkanen:2016jic}
T.~Tenkanen and V.~Vaskonen, \emph{{Reheating the Standard Model from a hidden
  sector}}, \href{https://doi.org/10.1103/PhysRevD.94.083516}{\emph{Phys. Rev.
  D} {\bfseries 94} (2016) 083516}
  [\href{https://arxiv.org/abs/1606.00192}{{\ttfamily 1606.00192}}].

\bibitem{Dror:2016rxc}
J.A.~Dror, E.~Kuflik and W.H.~Ng, \emph{{Codecaying Dark Matter}},
  \href{https://doi.org/10.1103/PhysRevLett.117.211801}{\emph{Phys. Rev. Lett.}
  {\bfseries 117} (2016) 211801}
  [\href{https://arxiv.org/abs/1607.03110}{{\ttfamily 1607.03110}}].

\bibitem{Berlin:2016gtr}
A.~Berlin, D.~Hooper and G.~Krnjaic, \emph{{Thermal Dark Matter From A Highly
  Decoupled Sector}},
  \href{https://doi.org/10.1103/PhysRevD.94.095019}{\emph{Phys. Rev. D}
  {\bfseries 94} (2016) 095019}
  [\href{https://arxiv.org/abs/1609.02555}{{\ttfamily 1609.02555}}].

\bibitem{DEramo:2017gpl}
F.~D'Eramo, N.~Fernandez and S.~Profumo, \emph{{When the Universe Expands Too
  Fast: Relentless Dark Matter}},
  \href{https://doi.org/10.1088/1475-7516/2017/05/012}{\emph{JCAP} {\bfseries
  05} (2017) 012} [\href{https://arxiv.org/abs/1703.04793}{{\ttfamily
  1703.04793}}].

\bibitem{Hamdan:2017psw}
S.~Hamdan and J.~Unwin, \emph{{Dark Matter Freeze-out During Matter
  Domination}}, \href{https://doi.org/10.1142/S021773231850181X}{\emph{Mod.
  Phys. Lett. A} {\bfseries 33} (2018) 1850181}
  [\href{https://arxiv.org/abs/1710.03758}{{\ttfamily 1710.03758}}].

\bibitem{Visinelli:2017qga}
L.~Visinelli, \emph{{(Non-)thermal production of WIMPs during kination}},
  \href{https://doi.org/10.3390/sym10110546}{\emph{Symmetry} {\bfseries 10}
  (2018) 546} [\href{https://arxiv.org/abs/1710.11006}{{\ttfamily
  1710.11006}}].

\bibitem{Dror:2017gjq}
J.A.~Dror, E.~Kuflik, B.~Melcher and S.~Watson, \emph{{Concentrated dark
  matter: Enhanced small-scale structure from codecaying dark matter}},
  \href{https://doi.org/10.1103/PhysRevD.97.063524}{\emph{Phys. Rev. D}
  {\bfseries 97} (2018) 063524}
  [\href{https://arxiv.org/abs/1711.04773}{{\ttfamily 1711.04773}}].

\bibitem{Drees:2017iod}
M.~Drees and F.~Hajkarim, \emph{{Dark Matter Production in an Early Matter
  Dominated Era}},
  \href{https://doi.org/10.1088/1475-7516/2018/02/057}{\emph{JCAP} {\bfseries
  02} (2018) 057} [\href{https://arxiv.org/abs/1711.05007}{{\ttfamily
  1711.05007}}].

\bibitem{DEramo:2017ecx}
F.~D'Eramo, N.~Fernandez and S.~Profumo, \emph{{Dark Matter Freeze-in
  Production in Fast-Expanding Universes}},
  \href{https://doi.org/10.1088/1475-7516/2018/02/046}{\emph{JCAP} {\bfseries
  02} (2018) 046} [\href{https://arxiv.org/abs/1712.07453}{{\ttfamily
  1712.07453}}].

\bibitem{Maity:2018dgy}
D.~Maity and P.~Saha, \emph{{Connecting CMB anisotropy and cold dark matter
  phenomenology via reheating}},
  \href{https://doi.org/10.1103/PhysRevD.98.103525}{\emph{Phys. Rev. D}
  {\bfseries 98} (2018) 103525}
  [\href{https://arxiv.org/abs/1801.03059}{{\ttfamily 1801.03059}}].

\bibitem{Bernal:2018ins}
N.~Bernal, C.~Cosme and T.~Tenkanen, \emph{{Phenomenology of Self-Interacting
  Dark Matter in a Matter-Dominated Universe}},
  \href{https://doi.org/10.1140/epjc/s10052-019-6608-8}{\emph{Eur. Phys. J. C}
  {\bfseries 79} (2019) 99} [\href{https://arxiv.org/abs/1803.08064}{{\ttfamily
  1803.08064}}].

\bibitem{Hardy:2018bph}
E.~Hardy, \emph{{Higgs portal dark matter in non-standard cosmological
  histories}}, \href{https://doi.org/10.1007/JHEP06(2018)043}{\emph{JHEP}
  {\bfseries 06} (2018) 043}
  [\href{https://arxiv.org/abs/1804.06783}{{\ttfamily 1804.06783}}].

\bibitem{Maity:2018exj}
D.~Maity and P.~Saha, \emph{{CMB constraints on dark matter phenomenology via
  reheating in Minimal plateau inflation}},
  \href{https://doi.org/10.1016/j.dark.2019.100317}{\emph{Phys. Dark Univ.}
  {\bfseries 25} (2019) 100317}
  [\href{https://arxiv.org/abs/1804.10115}{{\ttfamily 1804.10115}}].

\bibitem{Hambye:2018qjv}
T.~Hambye, A.~Strumia and D.~Teresi, \emph{{Super-cool Dark Matter}},
  \href{https://doi.org/10.1007/JHEP08(2018)188}{\emph{JHEP} {\bfseries 08}
  (2018) 188} [\href{https://arxiv.org/abs/1805.01473}{{\ttfamily
  1805.01473}}].

\bibitem{Bernal:2018kcw}
N.~Bernal, C.~Cosme, T.~Tenkanen and V.~Vaskonen, \emph{{Scalar singlet dark
  matter in non-standard cosmologies}},
  \href{https://doi.org/10.1140/epjc/s10052-019-6550-9}{\emph{Eur. Phys. J. C}
  {\bfseries 79} (2019) 30} [\href{https://arxiv.org/abs/1806.11122}{{\ttfamily
  1806.11122}}].

\bibitem{Arbey:2018uho}
A.~Arbey, J.~Ellis, F.~Mahmoudi and G.~Robbins, \emph{{Dark Matter Casts Light
  on the Early Universe}},
  \href{https://doi.org/10.1007/JHEP10(2018)132}{\emph{JHEP} {\bfseries 10}
  (2018) 132} [\href{https://arxiv.org/abs/1807.00554}{{\ttfamily
  1807.00554}}].

\bibitem{Drees:2018dsj}
M.~Drees and F.~Hajkarim, \emph{{Neutralino Dark Matter in Scenarios with Early
  Matter Domination}},
  \href{https://doi.org/10.1007/JHEP12(2018)042}{\emph{JHEP} {\bfseries 12}
  (2018) 042} [\href{https://arxiv.org/abs/1808.05706}{{\ttfamily
  1808.05706}}].

\bibitem{Betancur:2018xtj}
A.~Betancur and {\'O}.~Zapata, \emph{{Phenomenology of doublet-triplet
  fermionic dark matter in nonstandard cosmology and multicomponent dark
  sectors}}, \href{https://doi.org/10.1103/PhysRevD.98.095003}{\emph{Phys. Rev.
  D} {\bfseries 98} (2018) 095003}
  [\href{https://arxiv.org/abs/1809.04990}{{\ttfamily 1809.04990}}].

\bibitem{Maldonado:2019qmp}
C.~Maldonado and J.~Unwin, \emph{{Establishing the Dark Matter Relic Density in
  an Era of Particle Decays}},
  \href{https://doi.org/10.1088/1475-7516/2019/06/037}{\emph{JCAP} {\bfseries
  06} (2019) 037} [\href{https://arxiv.org/abs/1902.10746}{{\ttfamily
  1902.10746}}].

\bibitem{Poulin:2019omz}
A.~Poulin, \emph{{Dark matter freeze-out in modified cosmological scenarios}},
  \href{https://doi.org/10.1103/PhysRevD.100.043022}{\emph{Phys. Rev. D}
  {\bfseries 100} (2019) 043022}
  [\href{https://arxiv.org/abs/1905.03126}{{\ttfamily 1905.03126}}].

\bibitem{Tenkanen:2019cik}
T.~Tenkanen, \emph{{Standard model Higgs field and hidden sector cosmology}},
  \href{https://doi.org/10.1103/PhysRevD.100.083515}{\emph{Phys. Rev. D}
  {\bfseries 100} (2019) 083515}
  [\href{https://arxiv.org/abs/1905.11737}{{\ttfamily 1905.11737}}].

\bibitem{Han:2019vxi}
C.~Han, \emph{{Higgsino Dark Matter in a Non-Standard History of the
  Universe}}, \href{https://doi.org/10.1016/j.physletb.2019.134997}{\emph{Phys.
  Lett. B} {\bfseries 798} (2019) 134997}
  [\href{https://arxiv.org/abs/1907.09235}{{\ttfamily 1907.09235}}].

\bibitem{Bernal:2019mhf}
N.~Bernal, F.~Elahi, C.~Maldonado and J.~Unwin, \emph{{Ultraviolet Freeze-in
  and Non-Standard Cosmologies}},
  \href{https://doi.org/10.1088/1475-7516/2019/11/026}{\emph{JCAP} {\bfseries
  11} (2019) 026} [\href{https://arxiv.org/abs/1909.07992}{{\ttfamily
  1909.07992}}].

\bibitem{Chanda:2019xyl}
P.~Chanda, S.~Hamdan and J.~Unwin, \emph{{Reviving $Z$ and Higgs Mediated Dark
  Matter Models in Matter Dominated Freeze-out}},
  \href{https://doi.org/10.1088/1475-7516/2020/01/034}{\emph{JCAP} {\bfseries
  01} (2020) 034} [\href{https://arxiv.org/abs/1911.02616}{{\ttfamily
  1911.02616}}].

\bibitem{Mahanta:2019sfo}
D.~Mahanta and D.~Borah, \emph{{TeV Scale Leptogenesis with Dark Matter in
  Non-standard Cosmology}},
  \href{https://doi.org/10.1088/1475-7516/2020/04/032}{\emph{JCAP} {\bfseries
  04} (2020) 032} [\href{https://arxiv.org/abs/1912.09726}{{\ttfamily
  1912.09726}}].

\bibitem{Arcadi:2020aot}
G.~Arcadi, S.~Profumo, F.S.~Queiroz and C.~Siqueira, \emph{{Right-handed
  Neutrino Dark Matter, Neutrino Masses, and non-Standard Cosmology in a
  2HDM}}, \href{https://doi.org/10.1088/1475-7516/2020/12/030}{\emph{JCAP}
  {\bfseries 12} (2020) 030}
  [\href{https://arxiv.org/abs/2007.07920}{{\ttfamily 2007.07920}}].

\bibitem{Konar:2020vuu}
P.~Konar, A.~Mukherjee, A.K.~Saha and S.~Show, \emph{{A dark clue to seesaw and
  leptogenesis in a pseudo-Dirac singlet doublet scenario with (non)standard
  cosmology}}, \href{https://doi.org/10.1007/JHEP03(2021)044}{\emph{JHEP}
  {\bfseries 03} (2021) 044}
  [\href{https://arxiv.org/abs/2007.15608}{{\ttfamily 2007.15608}}].

\bibitem{Bhatia:2020itt}
D.~Bhatia and S.~Mukhopadhyay, \emph{{Unitarity limits on thermal dark matter
  in (non-)standard cosmologies}},
  \href{https://doi.org/10.1007/JHEP03(2021)133}{\emph{JHEP} {\bfseries 03}
  (2021) 133} [\href{https://arxiv.org/abs/2010.09762}{{\ttfamily
  2010.09762}}].

\bibitem{Arias:2020qty}
P.~Arias, D.~Karamitros and L.~Roszkowski, \emph{{Frozen-in fermionic singlet
  dark matter in non-standard cosmology with a decaying fluid}},
  \href{https://doi.org/10.1088/1475-7516/2021/05/041}{\emph{JCAP} {\bfseries
  05} (2021) 041} [\href{https://arxiv.org/abs/2012.07202}{{\ttfamily
  2012.07202}}].

\bibitem{Barman:2021ifu}
B.~Barman, P.~Ghosh, F.S.~Queiroz and A.K.~Saha, \emph{{Scalar multiplet dark
  matter in a fast expanding Universe: Resurrection of the desert region}},
  \href{https://doi.org/10.1103/PhysRevD.104.015040}{\emph{Phys. Rev. D}
  {\bfseries 104} (2021) 015040}
  [\href{https://arxiv.org/abs/2101.10175}{{\ttfamily 2101.10175}}].

\bibitem{Arcadi:2021doo}
G.~Arcadi, J.P.~Neto, F.S.~Queiroz and C.~Siqueira, \emph{{Roads for
  right-handed neutrino dark matter: Fast expansion, standard freeze-out, and
  early matter domination}},
  \href{https://doi.org/10.1103/PhysRevD.105.035016}{\emph{Phys. Rev. D}
  {\bfseries 105} (2022) 035016}
  [\href{https://arxiv.org/abs/2108.11398}{{\ttfamily 2108.11398}}].

\bibitem{Mahanta:2022gsi}
D.~Mahanta and D.~Borah, \emph{{WIMPy Leptogenesis in Non-Standard
  Cosmologies}},  \href{https://arxiv.org/abs/2208.11295}{{\ttfamily
  2208.11295}}.

\bibitem{Barrow:1982ei}
J.D.~Barrow, \emph{{Massive Particles as a Probe of the Early Universe}},
  \href{https://doi.org/10.1016/0550-3213(82)90233-4}{\emph{Nucl. Phys. B}
  {\bfseries 208} (1982) 501}.

\bibitem{Kamionkowski:1990ni}
M.~Kamionkowski and M.S.~Turner, \emph{{Thermal Relics: Do we Know their
  Abundances?}}, \href{https://doi.org/10.1103/PhysRevD.42.3310}{\emph{Phys.
  Rev. D} {\bfseries 42} (1990) 3310}.

\bibitem{McDonald:1989jd}
J.~McDonald, \emph{{{WIMP} Densities in Decaying Particle Dominated
  Cosmology}}, \href{https://doi.org/10.1103/PhysRevD.43.1063}{\emph{Phys. Rev.
  D} {\bfseries 43} (1991) 1063}.

\bibitem{Salati:2002md}
P.~Salati, \emph{{Quintessence and the relic density of neutralinos}},
  \href{https://doi.org/10.1016/j.physletb.2003.07.073}{\emph{Phys. Lett. B}
  {\bfseries 571} (2003) 121}
  [\href{https://arxiv.org/abs/astro-ph/0207396}{{\ttfamily
  astro-ph/0207396}}].

\bibitem{Comelli:2003cv}
D.~Comelli, M.~Pietroni and A.~Riotto, \emph{{Dark energy and dark matter}},
  \href{https://doi.org/10.1016/j.physletb.2003.05.006}{\emph{Phys. Lett. B}
  {\bfseries 571} (2003) 115}
  [\href{https://arxiv.org/abs/hep-ph/0302080}{{\ttfamily hep-ph/0302080}}].

\bibitem{Rosati:2003yw}
F.~Rosati, \emph{{Quintessential enhancement of dark matter abundance}},
  \href{https://doi.org/10.1016/j.physletb.2003.07.048}{\emph{Phys. Lett. B}
  {\bfseries 570} (2003) 5}
  [\href{https://arxiv.org/abs/hep-ph/0302159}{{\ttfamily hep-ph/0302159}}].

\bibitem{Pallis:2004yy}
C.~Pallis, \emph{{Massive particle decay and cold dark matter abundance}},
  \href{https://doi.org/10.1016/j.astropartphys.2004.05.006}{\emph{Astropart.
  Phys.} {\bfseries 21} (2004) 689}
  [\href{https://arxiv.org/abs/hep-ph/0402033}{{\ttfamily hep-ph/0402033}}].

\bibitem{Gelmini:2006pw}
G.B.~Gelmini and P.~Gondolo, \emph{{Neutralino with the right cold dark matter
  abundance in (almost) any supersymmetric model}},
  \href{https://doi.org/10.1103/PhysRevD.74.023510}{\emph{Phys. Rev. D}
  {\bfseries 74} (2006) 023510}
  [\href{https://arxiv.org/abs/hep-ph/0602230}{{\ttfamily hep-ph/0602230}}].

\bibitem{Gelmini:2006pq}
G.~Gelmini, P.~Gondolo, A.~Soldatenko and C.E.~Yaguna, \emph{{The Effect of a
  late decaying scalar on the neutralino relic density}},
  \href{https://doi.org/10.1103/PhysRevD.74.083514}{\emph{Phys. Rev. D}
  {\bfseries 74} (2006) 083514}
  [\href{https://arxiv.org/abs/hep-ph/0605016}{{\ttfamily hep-ph/0605016}}].

\bibitem{Arbey:2008kv}
A.~Arbey and F.~Mahmoudi, \emph{{SUSY constraints from relic density: High
  sensitivity to pre-BBN expansion rate}},
  \href{https://doi.org/10.1016/j.physletb.2008.09.032}{\emph{Phys. Lett. B}
  {\bfseries 669} (2008) 46} [\href{https://arxiv.org/abs/0803.0741}{{\ttfamily
  0803.0741}}].

\bibitem{Cohen:2008nb}
T.~Cohen, D.E.~Morrissey and A.~Pierce, \emph{{Changes in Dark Matter
  Properties After Freeze-Out}},
  \href{https://doi.org/10.1103/PhysRevD.78.111701}{\emph{Phys. Rev. D}
  {\bfseries 78} (2008) 111701}
  [\href{https://arxiv.org/abs/0808.3994}{{\ttfamily 0808.3994}}].

\bibitem{Arbey:2009gt}
A.~Arbey and F.~Mahmoudi, \emph{{SUSY Constraints, Relic Density, and Very
  Early Universe}}, \href{https://doi.org/10.1007/JHEP05(2010)051}{\emph{JHEP}
  {\bfseries 05} (2010) 051} [\href{https://arxiv.org/abs/0906.0368}{{\ttfamily
  0906.0368}}].

\bibitem{Giudice:2000ex}
G.F.~Giudice, E.W.~Kolb and A.~Riotto, \emph{{Largest temperature of the
  radiation era and its cosmological implications}},
  \href{https://doi.org/10.1103/PhysRevD.64.023508}{\emph{Phys. Rev. D}
  {\bfseries 64} (2001) 023508}
  [\href{https://arxiv.org/abs/hep-ph/0005123}{{\ttfamily hep-ph/0005123}}].

\bibitem{Fornengo:2002db}
N.~Fornengo, A.~Riotto and S.~Scopel, \emph{{Supersymmetric dark matter and the
  reheating temperature of the universe}},
  \href{https://doi.org/10.1103/PhysRevD.67.023514}{\emph{Phys. Rev. D}
  {\bfseries 67} (2003) 023514}
  [\href{https://arxiv.org/abs/hep-ph/0208072}{{\ttfamily hep-ph/0208072}}].

\bibitem{Drees:2006vh}
M.~Drees, H.~Iminniyaz and M.~Kakizaki, \emph{{Abundance of cosmological relics
  in low-temperature scenarios}},
  \href{https://doi.org/10.1103/PhysRevD.73.123502}{\emph{Phys. Rev. D}
  {\bfseries 73} (2006) 123502}
  [\href{https://arxiv.org/abs/hep-ph/0603165}{{\ttfamily hep-ph/0603165}}].

\bibitem{Roszkowski:2014lga}
L.~Roszkowski, S.~Trojanowski and K.~Turzy\'nski, \emph{{Neutralino and
  gravitino dark matter with low reheating temperature}},
  \href{https://doi.org/10.1007/JHEP11(2014)146}{\emph{JHEP} {\bfseries 11}
  (2014) 146} [\href{https://arxiv.org/abs/1406.0012}{{\ttfamily 1406.0012}}].

\bibitem{Davidson:2000dw}
S.~Davidson, M.~Losada and A.~Riotto, \emph{{A New perspective on
  baryogenesis}},
  \href{https://doi.org/10.1103/PhysRevLett.84.4284}{\emph{Phys. Rev. Lett.}
  {\bfseries 84} (2000) 4284}
  [\href{https://arxiv.org/abs/hep-ph/0001301}{{\ttfamily hep-ph/0001301}}].

\bibitem{Allahverdi:2010im}
R.~Allahverdi, B.~Dutta and K.~Sinha, \emph{{Baryogenesis and Late-Decaying
  Moduli}}, \href{https://doi.org/10.1103/PhysRevD.82.035004}{\emph{Phys. Rev.
  D} {\bfseries 82} (2010) 035004}
  [\href{https://arxiv.org/abs/1005.2804}{{\ttfamily 1005.2804}}].

\bibitem{Beniwal:2017eik}
A.~Beniwal, M.~Lewicki, J.D.~Wells, M.~White and A.G.~Williams,
  \emph{{Gravitational wave, collider and dark matter signals from a scalar
  singlet electroweak baryogenesis}},
  \href{https://doi.org/10.1007/JHEP08(2017)108}{\emph{JHEP} {\bfseries 08}
  (2017) 108} [\href{https://arxiv.org/abs/1702.06124}{{\ttfamily
  1702.06124}}].

\bibitem{Allahverdi:2017edd}
R.~Allahverdi, P.S.B.~Dev and B.~Dutta, \emph{{A simple testable model of
  baryon number violation: Baryogenesis, dark matter,
  neutron\textendash{}antineutron oscillation and collider signals}},
  \href{https://doi.org/10.1016/j.physletb.2018.02.019}{\emph{Phys. Lett. B}
  {\bfseries 779} (2018) 262}
  [\href{https://arxiv.org/abs/1712.02713}{{\ttfamily 1712.02713}}].

\bibitem{Bernal:2017zvx}
N.~Bernal and C.S.~Fong, \emph{{Hot Leptogenesis from Thermal Dark Matter}},
  \href{https://doi.org/10.1088/1475-7516/2017/10/042}{\emph{JCAP} {\bfseries
  10} (2017) 042} [\href{https://arxiv.org/abs/1707.02988}{{\ttfamily
  1707.02988}}].

\bibitem{Chen:2019etb}
S.-L.~Chen, A.~Dutta~Banik and Z.-K.~Liu, \emph{{Leptogenesis in fast expanding
  Universe}}, \href{https://doi.org/10.1088/1475-7516/2020/03/009}{\emph{JCAP}
  {\bfseries 03} (2020) 009}
  [\href{https://arxiv.org/abs/1912.07185}{{\ttfamily 1912.07185}}].

\bibitem{Chakraborty:2022gob}
M.~Chakraborty and S.~Roy, \emph{{Baryon asymmetry and lower bound on right
  handed neutrino mass in fast expanding Universe: an analytical approach}},
  \href{https://arxiv.org/abs/2208.04046}{{\ttfamily 2208.04046}}.

\bibitem{Assadullahi:2009nf}
H.~Assadullahi and D.~Wands, \emph{{Gravitational waves from an early matter
  era}}, \href{https://doi.org/10.1103/PhysRevD.79.083511}{\emph{Phys. Rev. D}
  {\bfseries 79} (2009) 083511}
  [\href{https://arxiv.org/abs/0901.0989}{{\ttfamily 0901.0989}}].

\bibitem{Durrer:2011bi}
R.~Durrer and J.~Hasenkamp, \emph{{Testing Superstring Theories with
  Gravitational Waves}},
  \href{https://doi.org/10.1103/PhysRevD.84.064027}{\emph{Phys. Rev. D}
  {\bfseries 84} (2011) 064027}
  [\href{https://arxiv.org/abs/1105.5283}{{\ttfamily 1105.5283}}].

\bibitem{Alabidi:2013lya}
L.~Alabidi, K.~Kohri, M.~Sasaki and Y.~Sendouda, \emph{{Observable induced
  gravitational waves from an early matter phase}},
  \href{https://doi.org/10.1088/1475-7516/2013/05/033}{\emph{JCAP} {\bfseries
  05} (2013) 033} [\href{https://arxiv.org/abs/1303.4519}{{\ttfamily
  1303.4519}}].

\bibitem{DEramo:2019tit}
F.~D'Eramo and K.~Schmitz, \emph{{Imprint of a scalar era on the primordial
  spectrum of gravitational waves}},
  \href{https://doi.org/10.1103/PhysRevResearch.1.013010}{\emph{Phys. Rev.
  Research.} {\bfseries 1} (2019) 013010}
  [\href{https://arxiv.org/abs/1904.07870}{{\ttfamily 1904.07870}}].

\bibitem{Bernal:2019lpc}
N.~Bernal and F.~Hajkarim, \emph{{Primordial Gravitational Waves in Nonstandard
  Cosmologies}}, \href{https://doi.org/10.1103/PhysRevD.100.063502}{\emph{Phys.
  Rev. D} {\bfseries 100} (2019) 063502}
  [\href{https://arxiv.org/abs/1905.10410}{{\ttfamily 1905.10410}}].

\bibitem{Figueroa:2019paj}
D.G.~Figueroa and E.H.~Tanin, \emph{{Ability of LIGO and LISA to probe the
  equation of state of the early Universe}},
  \href{https://doi.org/10.1088/1475-7516/2019/08/011}{\emph{JCAP} {\bfseries
  08} (2019) 011} [\href{https://arxiv.org/abs/1905.11960}{{\ttfamily
  1905.11960}}].

\bibitem{Bernal:2020ywq}
N.~Bernal, A.~Ghoshal, F.~Hajkarim and G.~Lambiase, \emph{{Primordial
  Gravitational Wave Signals in Modified Cosmologies}},
  \href{https://doi.org/10.1088/1475-7516/2020/11/051}{\emph{JCAP} {\bfseries
  11} (2020) 051} [\href{https://arxiv.org/abs/2008.04959}{{\ttfamily
  2008.04959}}].

\bibitem{Garcia:2022vwm}
M.A.G.~Garcia, M.~Pierre and S.~Verner, \emph{{Scalar Dark Matter Production
  from Preheating and Structure Formation Constraints}},
  \href{https://arxiv.org/abs/2206.08940}{{\ttfamily 2206.08940}}.

\bibitem{Kaneta:2022gug}
K.~Kaneta, S.M.~Lee and K.-y.~Oda, \emph{{Boltzmann or Bogoliubov? Approaches
  compared in gravitational particle production}},
  \href{https://doi.org/10.1088/1475-7516/2022/09/018}{\emph{JCAP} {\bfseries
  09} (2022) 018} [\href{https://arxiv.org/abs/2206.10929}{{\ttfamily
  2206.10929}}].

\bibitem{Kallosh:2013hoa}
R.~Kallosh and A.~Linde, \emph{{Universality Class in Conformal Inflation}},
  \href{https://doi.org/10.1088/1475-7516/2013/07/002}{\emph{JCAP} {\bfseries
  07} (2013) 002} [\href{https://arxiv.org/abs/1306.5220}{{\ttfamily
  1306.5220}}].

\bibitem{Kallosh:2013yoa}
R.~Kallosh, A.~Linde and D.~Roest, \emph{{Superconformal Inflationary
  $\alpha$-Attractors}},
  \href{https://doi.org/10.1007/JHEP11(2013)198}{\emph{JHEP} {\bfseries 11}
  (2013) 198} [\href{https://arxiv.org/abs/1311.0472}{{\ttfamily 1311.0472}}].

\bibitem{Starobinsky:1980te}
A.A.~Starobinsky, \emph{{A New Type of Isotropic Cosmological Models Without
  Singularity}},
  \href{https://doi.org/10.1016/0370-2693(80)90670-X}{\emph{Phys. Lett.}
  {\bfseries 91B} (1980) 99}.

\bibitem{Starobinsky:1981vz}
A.A.~Starobinsky, \emph{{Nonsingular Model of the Universe with the Quantum
  Gravitational de Sitter Stage and its Observational Consequences}},  in
  \emph{{Second Seminar on Quantum Gravity}}, pp.~103--128, 1, 1981.

\bibitem{Starobinsky:1983zz}
A.A.~Starobinsky, \emph{{The Perturbation Spectrum Evolving from a Nonsingular
  Initially De-Sitter Cosmology and the Microwave Background Anisotropy}},
  {\emph{Sov. Astron. Lett.} {\bfseries 9} (1983) 302}.

\bibitem{Kofman:1985aw}
L.~Kofman, A.D.~Linde and A.A.~Starobinsky, \emph{{Inflationary Universe
  Generated by the Combined Action of a Scalar Field and Gravitational Vacuum
  Polarization}},
  \href{https://doi.org/10.1016/0370-2693(85)90381-8}{\emph{Phys. Lett. B}
  {\bfseries 157} (1985) 361}.

\bibitem{Turner:1983he}
M.S.~Turner, \emph{{Coherent Scalar Field Oscillations in an Expanding
  Universe}}, \href{https://doi.org/10.1103/PhysRevD.28.1243}{\emph{Phys. Rev.
  D} {\bfseries 28} (1983) 1243}.

\bibitem{Garcia:2020eof}
M.A.G.~Garcia, K.~Kaneta, Y.~Mambrini and K.A.~Olive, \emph{{Reheating and
  Post-inflationary Production of Dark Matter}},
  \href{https://doi.org/10.1103/PhysRevD.101.123507}{\emph{Phys. Rev. D}
  {\bfseries 101} (2020) 123507}
  [\href{https://arxiv.org/abs/2004.08404}{{\ttfamily 2004.08404}}].

\bibitem{Garcia:2020wiy}
M.A.G.~Garcia, K.~Kaneta, Y.~Mambrini and K.A.~Olive, \emph{{Inflaton
  Oscillations and Post-Inflationary Reheating}},
  \href{https://doi.org/10.1088/1475-7516/2021/04/012}{\emph{JCAP} {\bfseries
  04} (2021) 012} [\href{https://arxiv.org/abs/2012.10756}{{\ttfamily
  2012.10756}}].

\bibitem{Co:2020xaf}
R.T.~Co, E.~Gonzalez and K.~Harigaya, \emph{{Increasing Temperature toward the
  Completion of Reheating}},
  \href{https://doi.org/10.1088/1475-7516/2020/11/038}{\emph{JCAP} {\bfseries
  11} (2020) 038} [\href{https://arxiv.org/abs/2007.04328}{{\ttfamily
  2007.04328}}].

\bibitem{Ahmed:2021fvt}
A.~Ahmed, B.~Grzadkowski and A.~Socha, \emph{{Implications of time-dependent
  inflaton decay on reheating and dark matter production}},
  \href{https://doi.org/10.1016/j.physletb.2022.137201}{\emph{Phys. Lett. B}
  {\bfseries 831} (2022) 137201}
  [\href{https://arxiv.org/abs/2111.06065}{{\ttfamily 2111.06065}}].

\bibitem{Barman:2022tzk}
B.~Barman, N.~Bernal, Y.~Xu and {\'O}.~Zapata, \emph{{Ultraviolet freeze-in
  with a time-dependent inflaton decay}},
  \href{https://doi.org/10.1088/1475-7516/2022/07/019}{\emph{JCAP} {\bfseries
  07} (2022) 019} [\href{https://arxiv.org/abs/2202.12906}{{\ttfamily
  2202.12906}}].

\bibitem{Ahmed:2022tfm}
A.~Ahmed, B.~Grzadkowski and A.~Socha, \emph{{Higgs boson induced reheating and
  ultraviolet frozen-in dark matter}},
  \href{https://arxiv.org/abs/2207.11218}{{\ttfamily 2207.11218}}.

\bibitem{Arias:2022qjt}
P.~Arias, N.~Bernal, J.K.~Osi\'nski and L.~Roszkowski, \emph{{Dark Matter
  Axions in the Early Universe with a Period of Increasing Temperature}},
  \href{https://arxiv.org/abs/2207.07677}{{\ttfamily 2207.07677}}.

\bibitem{Sarkar:1995dd}
S.~Sarkar, \emph{{Big bang nucleosynthesis and physics beyond the standard
  model}}, \href{https://doi.org/10.1088/0034-4885/59/12/001}{\emph{Rept. Prog.
  Phys.} {\bfseries 59} (1996) 1493}
  [\href{https://arxiv.org/abs/hep-ph/9602260}{{\ttfamily hep-ph/9602260}}].

\bibitem{Kawasaki:2000en}
M.~Kawasaki, K.~Kohri and N.~Sugiyama, \emph{{MeV scale reheating temperature
  and thermalization of neutrino background}},
  \href{https://doi.org/10.1103/PhysRevD.62.023506}{\emph{Phys. Rev. D}
  {\bfseries 62} (2000) 023506}
  [\href{https://arxiv.org/abs/astro-ph/0002127}{{\ttfamily
  astro-ph/0002127}}].

\bibitem{Hannestad:2004px}
S.~Hannestad, \emph{{What is the lowest possible reheating temperature?}},
  \href{https://doi.org/10.1103/PhysRevD.70.043506}{\emph{Phys. Rev. D}
  {\bfseries 70} (2004) 043506}
  [\href{https://arxiv.org/abs/astro-ph/0403291}{{\ttfamily
  astro-ph/0403291}}].

\bibitem{DeBernardis:2008zz}
F.~De~Bernardis, L.~Pagano and A.~Melchiorri, \emph{{New constraints on the
  reheating temperature of the universe after WMAP-5}},
  \href{https://doi.org/10.1016/j.astropartphys.2008.09.005}{\emph{Astropart.
  Phys.} {\bfseries 30} (2008) 192}.

\bibitem{deSalas:2015glj}
P.F.~de~Salas, M.~Lattanzi, G.~Mangano, G.~Miele, S.~Pastor and O.~Pisanti,
  \emph{{Bounds on very low reheating scenarios after Planck}},
  \href{https://doi.org/10.1103/PhysRevD.92.123534}{\emph{Phys. Rev. D}
  {\bfseries 92} (2015) 123534}
  [\href{https://arxiv.org/abs/1511.00672}{{\ttfamily 1511.00672}}].

\bibitem{Shtanov:1994ce}
Y.~Shtanov, J.H.~Traschen and R.H.~Brandenberger, \emph{{Universe reheating
  after inflation}},
  \href{https://doi.org/10.1103/PhysRevD.51.5438}{\emph{Phys. Rev. D}
  {\bfseries 51} (1995) 5438}
  [\href{https://arxiv.org/abs/hep-ph/9407247}{{\ttfamily hep-ph/9407247}}].

\bibitem{Ichikawa:2008ne}
K.~Ichikawa, T.~Suyama, T.~Takahashi and M.~Yamaguchi, \emph{{Primordial
  Curvature Fluctuation and Its Non-Gaussianity in Models with Modulated
  Reheating}}, \href{https://doi.org/10.1103/PhysRevD.78.063545}{\emph{Phys.
  Rev. D} {\bfseries 78} (2008) 063545}
  [\href{https://arxiv.org/abs/0807.3988}{{\ttfamily 0807.3988}}].

\bibitem{Lozanov:2016hid}
K.D.~Lozanov and M.A.~Amin, \emph{{Equation of State and Duration to Radiation
  Domination after Inflation}},
  \href{https://doi.org/10.1103/PhysRevLett.119.061301}{\emph{Phys. Rev. Lett.}
  {\bfseries 119} (2017) 061301}
  [\href{https://arxiv.org/abs/1608.01213}{{\ttfamily 1608.01213}}].

\bibitem{Maity:2018qhi}
D.~Maity and P.~Saha, \emph{{(P)reheating after minimal Plateau Inflation and
  constraints from CMB}},
  \href{https://doi.org/10.1088/1475-7516/2019/07/018}{\emph{JCAP} {\bfseries
  07} (2019) 018} [\href{https://arxiv.org/abs/1811.11173}{{\ttfamily
  1811.11173}}].

\bibitem{Saha:2020bis}
P.~Saha, S.~Anand and L.~Sriramkumar, \emph{{Accounting for the time evolution
  of the equation of state parameter during reheating}},
  \href{https://doi.org/10.1103/PhysRevD.102.103511}{\emph{Phys. Rev. D}
  {\bfseries 102} (2020) 103511}
  [\href{https://arxiv.org/abs/2005.01874}{{\ttfamily 2005.01874}}].

\bibitem{Antusch:2020iyq}
S.~Antusch, D.G.~Figueroa, K.~Marschall and F.~Torrenti, \emph{{Energy
  distribution and equation of state of the early Universe: matching the end of
  inflation and the onset of radiation domination}},
  \href{https://doi.org/10.1016/j.physletb.2020.135888}{\emph{Phys. Lett. B}
  {\bfseries 811} (2020) 135888}
  [\href{https://arxiv.org/abs/2005.07563}{{\ttfamily 2005.07563}}].

\bibitem{Easther:2010mr}
R.~Easther, R.~Flauger and J.B.~Gilmore, \emph{{Delayed Reheating and the
  Breakdown of Coherent Oscillations}},
  \href{https://doi.org/10.1088/1475-7516/2011/04/027}{\emph{JCAP} {\bfseries
  04} (2011) 027} [\href{https://arxiv.org/abs/1003.3011}{{\ttfamily
  1003.3011}}].

\bibitem{Freese:2017ace}
K.~Freese, E.I.~Sfakianakis, P.~Stengel and L.~Visinelli, \emph{{The Higgs
  Boson can delay Reheating after Inflation}},
  \href{https://doi.org/10.1088/1475-7516/2018/05/067}{\emph{JCAP} {\bfseries
  05} (2018) 067} [\href{https://arxiv.org/abs/1712.03791}{{\ttfamily
  1712.03791}}].

\bibitem{Harigaya:2014waa}
K.~Harigaya, M.~Kawasaki, K.~Mukaida and M.~Yamada, \emph{{Dark Matter
  Production in Late Time Reheating}},
  \href{https://doi.org/10.1103/PhysRevD.89.083532}{\emph{Phys. Rev. D}
  {\bfseries 89} (2014) 083532}
  [\href{https://arxiv.org/abs/1402.2846}{{\ttfamily 1402.2846}}].

\bibitem{Harigaya:2019tzu}
K.~Harigaya, K.~Mukaida and M.~Yamada, \emph{{Dark Matter Production during the
  Thermalization Era}},
  \href{https://doi.org/10.1007/JHEP07(2019)059}{\emph{JHEP} {\bfseries 07}
  (2019) 059} [\href{https://arxiv.org/abs/1901.11027}{{\ttfamily
  1901.11027}}].

\bibitem{Drees:2021lbm}
M.~Drees and B.~Najjari, \emph{{Energy spectrum of thermalizing high energy
  decay products in the early universe}},
  \href{https://doi.org/10.1088/1475-7516/2021/10/009}{\emph{JCAP} {\bfseries
  10} (2021) 009} [\href{https://arxiv.org/abs/2105.01935}{{\ttfamily
  2105.01935}}].

\bibitem{Drees:2022vvn}
M.~Drees and B.~Najjari, \emph{{Multi-Species Thermalization Cascade of
  Energetic Particles in the Early Universe}},
  \href{https://arxiv.org/abs/2205.07741}{{\ttfamily 2205.07741}}.

\bibitem{Mukaida:2022bbo}
K.~Mukaida and M.~Yamada, \emph{{Cascades of high-energy SM particles in the
  primordial thermal plasma}},
  \href{https://arxiv.org/abs/2208.11708}{{\ttfamily 2208.11708}}.

\bibitem{ParticleDataGroup:2020ssz}
{\scshape Particle Data Group} collaboration, \emph{{Review of Particle
  Physics}}, \href{https://doi.org/10.1093/ptep/ptaa104}{\emph{PTEP} {\bfseries
  2020} (2020) 083C01}.

\bibitem{Leane:2018kjk}
R.K.~Leane, T.R.~Slatyer, J.F.~Beacom and K.C.Y.~Ng, \emph{{GeV-scale thermal
  WIMPs: Not even slightly ruled out}},
  \href{https://doi.org/10.1103/PhysRevD.98.023016}{\emph{Phys. Rev. D}
  {\bfseries 98} (2018) 023016}
  [\href{https://arxiv.org/abs/1805.10305}{{\ttfamily 1805.10305}}].

\bibitem{Hess:2021cdp}
{\scshape Hess, HAWC, VERITAS, MAGIC, H.E.S.S., Fermi-LAT} collaboration,
  \emph{{Combined dark matter searches towards dwarf spheroidal galaxies with
  Fermi-LAT, HAWC, H.E.S.S., MAGIC, and VERITAS}},
  \href{https://doi.org/10.22323/1.395.0528}{\emph{PoS} {\bfseries ICRC2021}
  (2021) 528} [\href{https://arxiv.org/abs/2108.13646}{{\ttfamily
  2108.13646}}].

\bibitem{Bergstrom:2013jra}
L.~Bergstrom, T.~Bringmann, I.~Cholis, D.~Hooper and C.~Weniger, \emph{{New
  Limits on Dark Matter Annihilation from AMS Cosmic Ray Positron Data}},
  \href{https://doi.org/10.1103/PhysRevLett.111.171101}{\emph{Phys. Rev. Lett.}
  {\bfseries 111} (2013) 171101}
  [\href{https://arxiv.org/abs/1306.3983}{{\ttfamily 1306.3983}}].

\bibitem{Calore:2022stf}
F.~Calore, M.~Cirelli, L.~Derome, Y.~Genolini, D.~Maurin, P.~Salati et~al.,
  \emph{{AMS-02 antiprotons and dark matter: Trimmed hints and robust bounds}},
  \href{https://doi.org/10.21468/SciPostPhys.12.5.163}{\emph{SciPost Phys.}
  {\bfseries 12} (2022) 163}
  [\href{https://arxiv.org/abs/2202.03076}{{\ttfamily 2202.03076}}].

\bibitem{Zemp:2008gw}
M.~Zemp, J.~Diemand, M.~Kuhlen, P.~Madau, B.~Moore, D.~Potter et~al.,
  \emph{{The Graininess of Dark Matter Haloes}},
  \href{https://doi.org/10.1111/j.1365-2966.2008.14361.x}{\emph{Mon. Not. Roy.
  Astron. Soc.} {\bfseries 394} (2009) 641}
  [\href{https://arxiv.org/abs/0812.2033}{{\ttfamily 0812.2033}}].

\bibitem{Pato:2010yq}
M.~Pato, O.~Agertz, G.~Bertone, B.~Moore and R.~Teyssier, \emph{{Systematic
  uncertainties in the determination of the local dark matter density}},
  \href{https://doi.org/10.1103/PhysRevD.82.023531}{\emph{Phys. Rev. D}
  {\bfseries 82} (2010) 023531}
  [\href{https://arxiv.org/abs/1006.1322}{{\ttfamily 1006.1322}}].

\bibitem{Bernal:2014mmt}
N.~Bernal, J.E.~Forero-Romero, R.~Garani and S.~Palomares-Ruiz,
  \emph{{Systematic uncertainties from halo asphericity in dark matter
  searches}}, \href{https://doi.org/10.1088/1475-7516/2014/09/004}{\emph{JCAP}
  {\bfseries 09} (2014) 004} [\href{https://arxiv.org/abs/1405.6240}{{\ttfamily
  1405.6240}}].

\bibitem{Boudaud:2014dta}
M.~Boudaud et~al., \emph{{A new look at the cosmic ray positron fraction}},
  \href{https://doi.org/10.1051/0004-6361/201425197}{\emph{Astron. Astrophys.}
  {\bfseries 575} (2015) A67}
  [\href{https://arxiv.org/abs/1410.3799}{{\ttfamily 1410.3799}}].

\bibitem{Bernal:2015oyn}
N.~Bernal, J.E.~Forero-Romero, R.~Garani and S.~Palomares-Ruiz,
  \emph{{Systematic uncertainties from halo asphericity in dark matter
  searches}},
  \href{https://doi.org/10.1016/j.nuclphysbps.2015.10.129}{\emph{Nucl. Part.
  Phys. Proc.} {\bfseries 267-269} (2015) 345}.

\bibitem{Bernal:2016guq}
N.~Bernal, L.~Necib and T.R.~Slatyer, \emph{{Spherical Cows in Dark Matter
  Indirect Detection}},
  \href{https://doi.org/10.1088/1475-7516/2016/12/030}{\emph{JCAP} {\bfseries
  12} (2016) 030} [\href{https://arxiv.org/abs/1606.00433}{{\ttfamily
  1606.00433}}].

\bibitem{Benito:2016kyp}
M.~Benito, N.~Bernal, N.~Bozorgnia, F.~Calore and F.~Iocco, \emph{{Particle
  Dark Matter Constraints: the Effect of Galactic Uncertainties}},
  \href{https://doi.org/10.1088/1475-7516/2017/02/007}{\emph{JCAP} {\bfseries
  02} (2017) 007} [\href{https://arxiv.org/abs/1612.02010}{{\ttfamily
  1612.02010}}].

\bibitem{CTA:2020qlo}
{\scshape CTA} collaboration, \emph{{Sensitivity of the Cherenkov Telescope
  Array to a dark matter signal from the Galactic centre}},
  \href{https://doi.org/10.1088/1475-7516/2021/01/057}{\emph{JCAP} {\bfseries
  01} (2021) 057} [\href{https://arxiv.org/abs/2007.16129}{{\ttfamily
  2007.16129}}].

\end{thebibliography}\endgroup

\end{document}